\documentclass[journal]{IEEEtran}

\usepackage{amssymb}
\usepackage{cite}
\usepackage{amsmath}
\usepackage{xcolor}
\usepackage{multirow}
\usepackage{float}
\usepackage{colortbl}
\usepackage{booktabs}
\usepackage{array}
\usepackage{mathtools, cuted}
\usepackage{makecell}
\usepackage{amsmath}%
\usepackage{enumitem}
\usepackage{dblfloatfix}
\usepackage{graphicx}

\definecolor{mycolor}{rgb}{.949,.949,.949}

\usepackage{hyperref} 
\hypersetup{
    colorlinks=true,       
    linkcolor=blue,        
    citecolor=blue,        
    urlcolor=blue,         
    pdfborder={0 0 0}      
}

\usepackage{algorithm}
\usepackage{algorithmic}

\usepackage{tikz}

\usepackage{comment}

\begin{document}
\bstctlcite{IEEEexample:BSTcontrol}

\title{Revisiting the Effect of Grid-Following Converter on Frequency Dynamics - Part I: Center of Inertia}
\vspace{-0.5em}
\author{Jiahao~Liu,~\IEEEmembership{Student Member,~IEEE}, Cheng~Wang,~\IEEEmembership{Senior Member,~IEEE}, Tianshu~Bi,~\IEEEmembership{Fellow,~IEEE}
\vspace{-3.27em}
}

\maketitle

\begin{abstract}
    Understanding the impact of grid-following (GFL) converters on system frequency dynamics is crucial, from both the center of inertia (COI) and frequency spatial variation perspectives. Part I of this series clarifies the mechanisms by which GFLs influence COI frequency dynamics. A multi-generator model of the power system with GFLs is developed, incorporating the local dynamics of GFLs and their interaction with synchronous generators via virtual tie lines. By aggregating the multi-generator model into the COI frame, the interaction between the COI frequency and the equivalent frequency of GFLs is revealed. The equivalent inertia and other components at the GFL side, determined by control parameters and operating conditions, support the COI through virtual tying power. Simulation validates the accuracy of the proposed modeling and demonstrates that the impact of GFLs on COI frequency is relatively weak. The equivalent inertia and other components of GFLs still significantly influence COI frequency dynamics, with their effects being both time-variable and adjustable.
\end{abstract}
\vspace{-0.5em}
\begin{IEEEkeywords}
	Frequency dynamics, grid-following, center of inertia, equivalent inertia. 
\end{IEEEkeywords}

\vspace{-1.3em} 
\section{Introduction\label{Sec:Introd}}
\vspace{-0.3em} 

\IEEEPARstart{D}URING the past decades, power system frequency dynamics is dominated by synchronous generators (SGs). At the generator level, SGs provide inertia response and primary frequency regulation \cite{KYan23Review}. At the system level, all SGs are naturally interconnected through the synchronizing torque, and the rotor frequency of SGs serving as system boundary defines the network node frequency \cite{FMilano17}. This clear influence of SGs impacting frequency dynamics facilitates both center of inertia (COI) and spatial frequency analyses \cite{FMilano18Review}.

The inverter-based resources (IBRs), predominantly grid-following (GFL) converters, introduce significant changes to frequency dynamics from both COI and spatial perspectives \cite{XChen24}. Unlike SGs, GFLs are passively synchronized through a phase-locked loop (PLL), and their local dynamics are shaped by the combined effects of various artificial control modules, such as active power control and PLL \cite{XHe21}. Consequently, comprehensively understanding the influence of GFLs on system frequency dynamics is critical for supporting frequency stability analysis and control in modern power systems.

However, the fundamentally different dynamics of GFLs compared to SGs present significant challenges. The generally accepted view is that GFLs prioritize maintaining stable output active power and do not actively contribute to frequency support, potentially exacerbating system frequency stability issues \cite{NHatziargyriou21, FMilano18Review}. Beyond this general understanding, there is no unified consensus on their detailed impact.

Let us first investigate the effect of GFLs on COI frequency dynamics, as the COI aspect serves as the foundation for frequency stability analysis. The existing perspectives on this topic can be categorized into three primary approaches.

The first is the \textbf{over-simplified approach}. All detailed dynamics in the GFL, such as active power control and PLL, are disregarded. Thus, GFLs without frequency regulation are treated as constant power sources during disturbances, contributing nothing to system frequency dynamics \cite{LXiong22}. GFLs with additional frequency regulation are modeled as generators possessing ideal inertia and damping capabilities \cite{KYan23}. Quantitatively, the traditional system frequency response (SFR) model, which describes COI dynamics in SG-dominated systems, can still be used to analyze GFLs \cite{KYan23}. Following this approach, many studies emerge, including setting system-wide total inertia and damping through unit commitment \cite{LBadesa19, ZZhang22, KLi24}, online allocating virtual inertia \cite{BShe24, ZPeng24}, and implementing real-time frequency control that incorporates GFLs \cite{TBaskarad23}. However, the high level of simplification inherent in this approach limits its applicability in future systems with higher levels of IBRs.

The second is the \textbf{equivalent governor approach}. Detailed dynamics within GFLs are incorporated. For instance, the mechanical rotor speed of wind-type GFLs decreases during virtual frequency regulation, reducing power support and causing a frequency drop during the recovery period \cite{CZhao23}. The maximum power point tracking (MPPT) logic, which determines the operating status of GFLs, also influences their virtual frequency regulation capabilities \cite{YYuan23}. These dynamics can be quantitatively represented by a transfer function that takes system frequency as the input and outputs active power, resembling the behavior of the speed governor in SGs. An enhanced SFR model can be achieved by incorporating these transfer functions. Examples of SFR models include those that account for detailed GFL dynamics, such as MPPT logic \cite{YYuan23}, mechanical rotor dynamics in wind systems \cite{CZhao23, YZhang24}, control systems for PV \cite{TBaskarad21}, and comprehensive models integrating all these modules \cite{JYu24, JHuang22, XZhang24}. However, because this approach uses system frequency as its input, it is limited to GFLs equipped with a virtual frequency regulation function. Moreover, the most critical module, the PLL, cannot be effectively captured within this governor-based framework.

The third is the \textbf{equivalent-rotor-motion approach}. This approach considers all internal dynamics of the GFL, including the PLL, and analyzes GFLs similarly to the swing equation of SGs, where active power serves as the input and electromotive force (EMF) angle acts as the output. The equivalent inertia and damping of GFLs are derived and compared to those of SGs to evaluate the impact of GFLs on system frequency dynamics. Based on the definition of the virtual EMF for GFLs, this approach can be further categorized into the internal-voltage-based approach, as presented in \cite{HYuan17, RFu22}, and the current-based approach, as outlined in \cite{MZhang18}. However, a notable limitation arises from the definition of the virtual EMF. For example, in \cite{HYuan17}, the virtual internal voltage is defined behind the terminal voltage and the converter filter, causing it to fail as a proper state variable, which contrasts with the role of SG's EMF. Additionally, the model accuracy heavily relies on filter parameters, which can introduce errors. Beyond this, the most critical limitation is that system-wide interactions remain unclear due to the fundamentally different synchronization mechanisms of SGs and GFLs \cite{YZhou22, CHe22}. It remains debatable whether the derived equivalent inertia of GFLs contributes to COI frequency in the same way as SGs.

Considering the ambiguity in GFL research, it is essential to revisit how SGs influence COI frequency dynamics. Two key aspects must be addressed \cite{PMAnderson03}. The first is the power-frequency response at the generator side, which is the primary focus of the above works aiming to accurately model GFLs. The second is the aggregation of all generators into the COI. This aggregation is feasible in SG-dominated systems due to the synchronizing torque connecting the SGs like a spring \cite{CJTavora72}. This torque ensures that the active power at both ends of the spring is equal, allowing all SGs to be aggregated into the COI frame, with each SG's inertia contribution being uniform.

In Part I of this series, the impact of GFLs on COI frequency is thoroughly examined. It addresses both the accurate modeling at the generator side and, for the first time, investigates how GFLs and SGs are aggregated into the COI frame. Compared to existing methods, this work makes three key contributions:

(1) The multi-generator representation of power systems containing both SGs and GFLs is developed to support COI frequency mechanism analysis. On the generator side, GFL dynamics are modeled in strict alignment with the swing equation of SGs, where the equivalent EMF of GFLs serves as a state variable. On the network side, the interaction between the state variables of SGs and GFLs is modeled using the concept of tying power between them.

(2) It is identified that the connecting power between SGs is equal in magnitude, making SGs interconnected like a spring-mass system. However, the connecting power between SGs and GFLs does not follow this pattern. Thus, only SGs form the system COI, while the equivalent frequency of GFLs interacts with the SG COI frequency through a virtual tie line.

(3) The effect of GFLs on COI frequency dynamics via connecting power is quantitatively analyzed. The connection between SGs and GFLs is weak for both paths driven by the angle and amplitude of the interface state variable. GFLs affect the COI frequency pattern with their equivalent inertia and equivalent governor, which are determined by their control parameters and operating states.

\vspace{-1.3em} 
\section{Problem Formulation\label{Sec:ProbForm}}
\vspace{-0.3em} 

\vspace{-0.2em} 
\subsection{COI Frequency Dynamics\label{SubSec:TraCoi}}
\vspace{-0.3em} 

\begin{figure}[!t]
	\centering
	\includegraphics[width=0.46\textwidth]{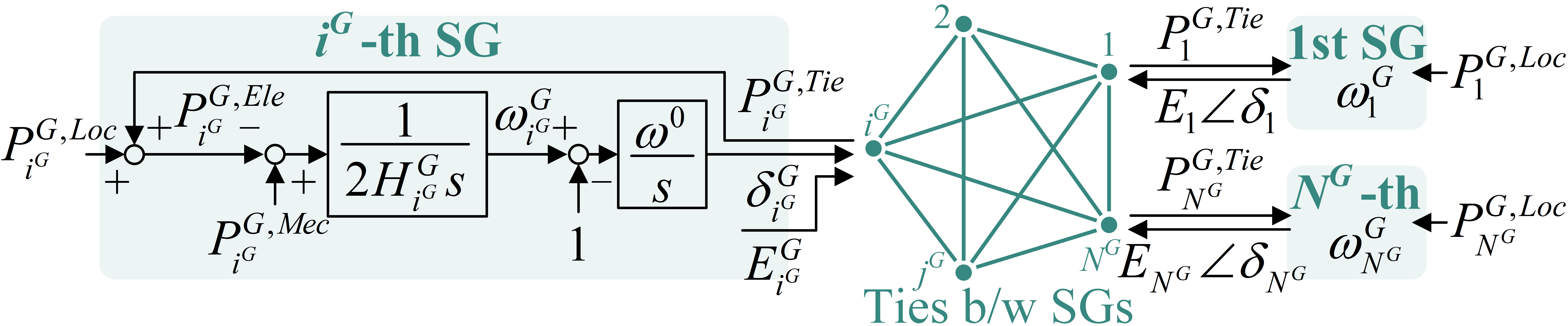}
    \vspace{-1.0em}
	\caption{Multi-generator dynamics in traditional power systems.}
    \vspace{-1.8em}
	\label{fig:mmCOISGs}
\end{figure}

The COI representation of frequency dynamics originates from the elegant structure within the system's multi-generator model as shown in Fig. \ref{fig:mmCOISGs} \cite{CJTavora72}. The core in Fig. \ref{fig:mmCOISGs} is the EMF of the SG's second-order model, which serves as the interface between the SG rotor dynamics and the power network:
\vspace{-0.5em} 
\begin{equation}\label{eq:DefSS}
    \text{interface state variable: } E^{G} \angle \delta^{G},
    \vspace{-0.7em} 
\end{equation}
where $E^{G}$ and $\delta^{G}$ are amplitude and phase angle components, respectively. Note that the EMF amplitude, $E^{G}$, is constant.

On the rotor dynamics side, the swing equation governs the dynamics of the phase angle of the interface state variable, $\delta^{G}$, as follows:
\begin{subequations}\label{eq:GeneSide}
\vspace{-0.5em} 
\begin{equation}\label{eq:GeneSide_H}
    2 H_{i^{G}}^{G} \frac{\mathrm{d} \omega_{i^{G}}^{G}}{\mathrm{d} t} = P_{i^{G}}^{G,Mec} - P_{i^{G}}^{G,Ele},
    \vspace{-0.7em} 
\end{equation}
\vspace{-0.5em} 
\begin{equation}\label{eq:GeneSide_d}
    \frac{\mathrm{d} \delta_{i^{G}}^{G}}{\mathrm{d} t} = \omega^{0} \left( \omega_{i^{G}}^{G} - 1 \right),
    \vspace{-0.7em} 
\end{equation}
\end{subequations}
where $i^{G}$ is the index for SGs; $\omega^{G}$ represents the rotor frequency; $P^{G,Ele}$ and $P^{G,Mec}$ denote the electromagenitic power and mechanical power, respectilvely; $H^{G}$ is the generator inertia constant; and $\omega^{0}$ is the nominal frequency.

On the network side, the generator active power $P^{G,Ele}$ is influenced by the differences in the EMF angles, $\delta^{G}$, between generators. The power transferred through a equivalent tie line, $P^{G,Tie}$, represents this relationship, as indicated by the ``tie b/w SGs'' in Fig. \ref{fig:mmCOISGs}. Together with the local power $P^{G,Loc}$ at the generator terminal, $P^{G,Ele}$ can be expressed as:
\begin{subequations}\label{eq:NetwSide}
\vspace{-0.5em} 
\begin{equation}\label{eq:NetwSide_E}
    P_{i^{G}}^{G,Ele} = P_{i^{G}}^{G,Loc} + P_{i^{G}}^{G,Tie} ,
    \vspace{-0.7em} 
\end{equation}
\vspace{-0.5em} 
\begin{equation}\label{eq:NetwSide_Tie}
    P_{i^{G}}^{G,Tie} = \sum_{\substack{\forall j^{G} \in N^{G} \\ j^{G} \neq i^{G}}} E_{i^{G}}^{G} E_{j^{G}}^{G} B_{i^{G},j^{G}}^{GG} \sin\left( \delta_{i^{G}}^{G} - \delta_{j^{G}}^{G}\right) ,
    \vspace{-0.7em} 
\end{equation}
\end{subequations}
where $j^{G}$ is also an index for SGs, and $N^{G}$ is the total number of SGs; $B^{GG}$ denotes the imaginary component of the equivalent admittance of equivalent ties.

\begin{figure}[!t]
	\centering
	\includegraphics[width=0.27\textwidth]{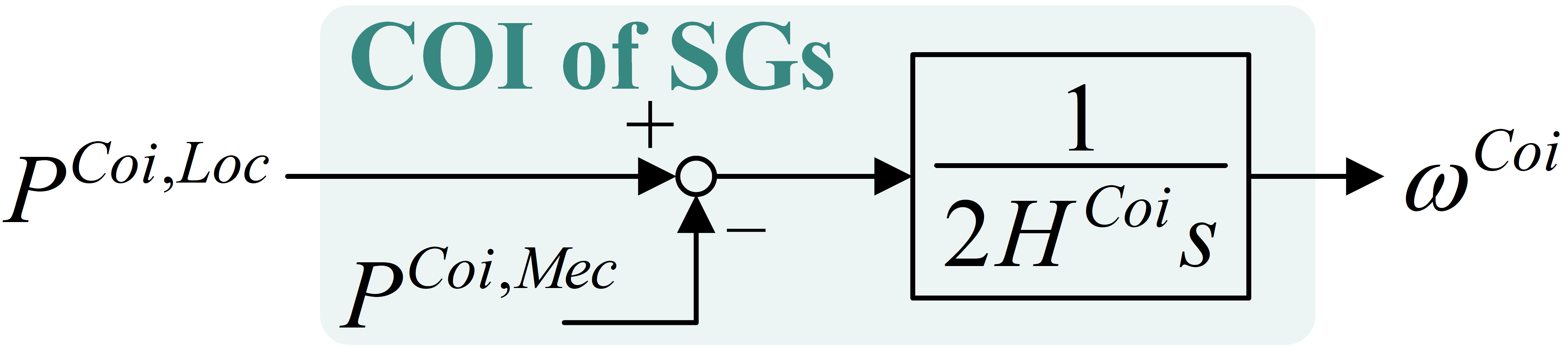}
    \vspace{-1.0em}
	\caption{COI frequency dynamics in traditional power systems.}
    \vspace{-2.0em}
	\label{fig:COISGs}
\end{figure}

The multi-generator system described by (\ref{eq:GeneSide}) and (\ref{eq:NetwSide}) satisfies the following conditions \cite{CJTavora72}: (i) the generator follows the inertia motion law; (ii) since $B^{GG}$ is symmetric, with $B_{i^{G},j^{G}}^{GG} = B_{j^{G},i^{G}}^{GG}$, the tying powers between any pair of generators like a spring, having equal magnitudes but opposite directions, that is $E_{i^{G}}^{G} E_{j^{G}}^{G} B_{i^{G},j^{G}}^{GG} \sin\left( \delta_{i^{G}}^{G} - \delta_{j^{G}}^{G}\right) + E_{j^{G}}^{G} E_{i^{G}}^{G} B_{j^{G},i^{G}}^{GG} \sin\left( \delta_{j^{G}}^{G} - \delta_{i^{G}}^{G}\right) = 0$. As a result, the multi-generator system can be aggregated into the COI frame by eliminating the tying power, as shown in Fig. \ref{fig:COISGs} with equations:
\vspace{-0.5em} 
\begin{equation}\label{eq:TradCOI}
    2 H^{Coi} \frac{\mathrm{d} \omega^{Coi}}{\mathrm{d} t} = P^{Coi,Mec} - P^{Coi,Loc},
    \vspace{-0.7em} 
\end{equation}
where $P^{Coi,Loc}$ and $P^{Coi,Mec}$ are the total local power and total mechanical power; the expressions for COI frequency $\omega^{Coi}$ and inertia $H^{Coi}$ are defined as:
\begin{subequations}\label{eq:TradCOIfH}
\vspace{-0.5em} 
\begin{equation}\label{eq:TradCOIfH_f}
    \omega^{Coi} = \sum_{\forall i^{G} \in N^{G}} \frac{S_{i^{G}}^{G} H_{i^{G}}^{G} \omega_{i^{G}}^{G}}{S_{i^{G}}^{G} H_{i^{G}}^{G}},
    \vspace{-0.7em} 
\end{equation}
\vspace{-0.5em} 
\begin{equation}\label{eq:TradCOIfH_H}
    H^{Coi} = \sum_{\forall i^{G} \in N^{G}} \frac{S_{i^{G}}^{G} H_{i^{G}}^{G}}{S_{i^{G}}^{G}},
    \vspace{-0.7em} 
\end{equation}
\end{subequations}
where $S^{G}$ represents the SG nominal power.

Using (\ref{eq:TradCOI}) and (\ref{eq:TradCOIfH}), or Fig. \ref{fig:COISGs}, the influence of SGs on the system average frequency can be analyzed. This includes examining the impact of inertia and governor on key metrics such as the COI maximum RoCoF and frequency nadir.

\vspace{-0.9em} 
\subsection{Tasks for GFL Study\label{SubSec:ProCoi}}
\vspace{-0.3em} 

First, the model of GFLs should be clarified. Power sources such as wind, PV, or battery storage provide the input power $P^{F,In}$. This power is then processed through a DC boost and filter to supply active power $P^{F,Ele}$ and reactive power $Q^{F}$ to the grid. At the PLL level, the PLL angle $\theta^{Pll}$ is generated by tracking the phase angle $\theta^{F,U}$ of the terminal voltage. At the control level, the outer controllers stabilize the DC voltage $U^{Dc}$ and regulate $Q^{F}$ or the terminal voltage amplitude $U^{F}$ using d-axis and q-axis PI controllers, respectively. The inner controllers adjust the d-axis current $i_{d}^{P}$ and q-axis current $i_{q}^{P}$. The superscript $\{\cdot\}^{P}$ denotes variables in the PLL angle frame, and $\{\cdot\}^{*}$ represents command values.

\begin{figure}[!t]
	\centering
	\includegraphics[width=0.46\textwidth]{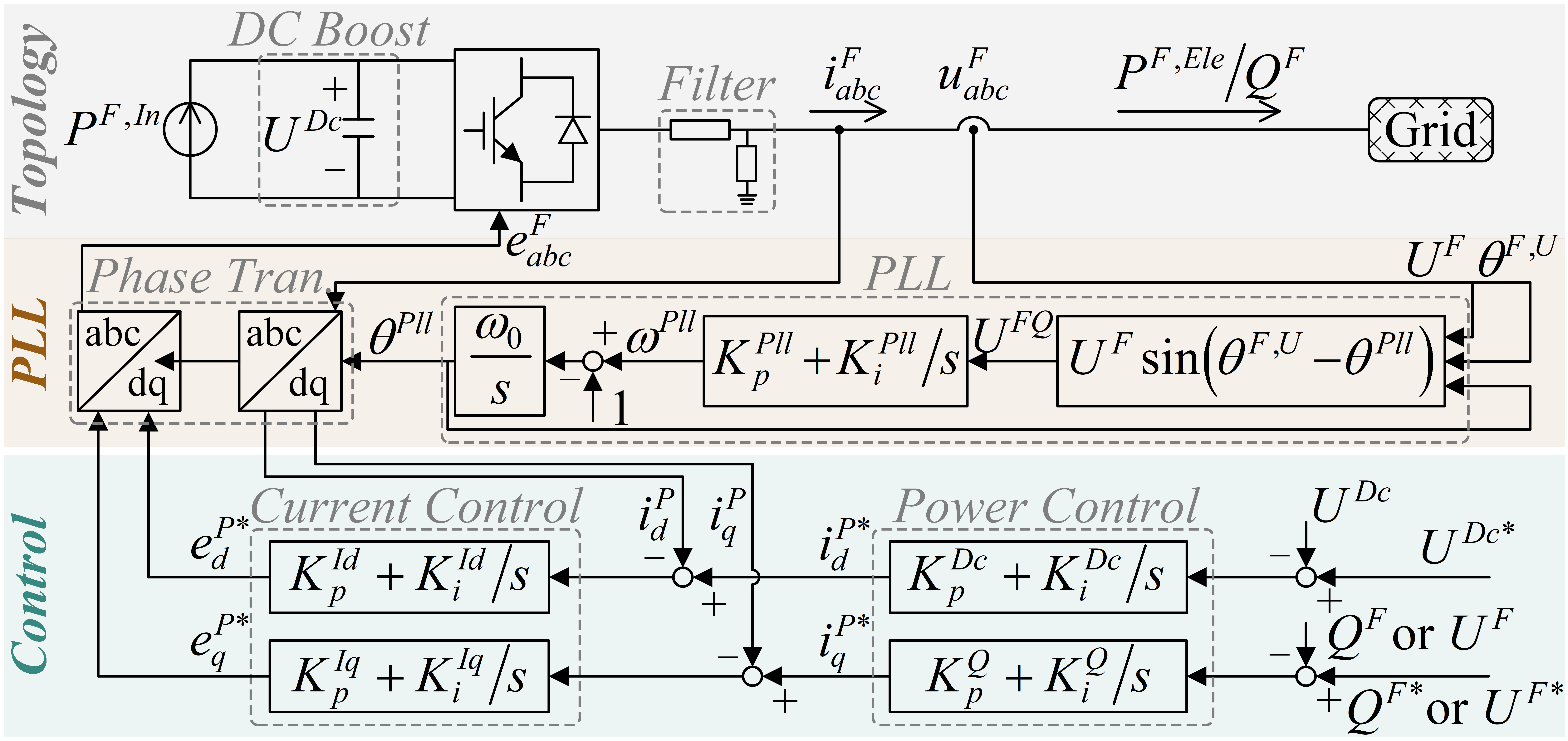}
    \vspace{-1.0em}
	\caption{Model of the generalized GFL-based generator.}
    \vspace{-1.5em}  
	\label{fig:GFLCont}
\end{figure}

Note that virtual frequency regulation and voltage droop control functions can be implemented by introducing additional commands based on $U^{DC*}$, $Q^{F*}$, and $U^{F*}$ \cite{MZhang18}. However, these are not considered in the main text of this paper because their inclusion does not impact the derivation presented in the following sections. The derivation under various additional control logics, such as virtual inertia and fast frequency response \cite{MDreidy17}, can be supplemented easily.

By analogy to the steps in Section \ref{SubSec:TraCoi}, the process for analyzing the effect of GFLs on COI frequency dynamics can be outlined. The main steps include:
\vspace{-0.2em} 
\begin{itemize}
    \item \textbf{Modeling multi-generator dynamics}: The first step is to include GFL converters in the system multi-generator representation. This involves three tasks: 1) identifying the GFL state variables interfaced with the power grid; 2) similar to the rotor dynamics (\ref{eq:GeneSide}) of SGs, characterizing the dynamics of GFL interface state variables with equivalent inertia; and 3) similar to (\ref{eq:NetwSide}), expressing the active power of SGs or GFLs as a function of the SG EMF and the GFL interfacing state variable.
    \item \textbf{Transferring to COI frequency dynamics and analyzing the effect}: The multi-generator modeling that includes GFLs can be aggregated into COI dynamics similar to (\ref{eq:TradCOI}). The mechanisms by which GFLs impact COI frequency dynamics can then be analyzed, such as the role of inertia contributed by the GFL.
\end{itemize}
\vspace{-0.0em}  

\vspace{-1.3em} 
\section{Multi-Generator Modeling of Frequency Dynamics Containing GFLs\label{Sec:MGmodel}}
\vspace{-0.3em} 

The interface state variable of GFL is defined in Section \ref{SubSec:SSPart}. Sections \ref{SubSec:GenePart} and \ref{SubSec:NetPart} provide modeling of the GFL dynamics and the network equations, respectively.

\vspace{-1.3em}  
\subsection{Definition of GFL Interface State Variable\label{SubSec:SSPart}}
\vspace{-0.3em} 

Considering the timescale of power system frequency dynamics, the following assumptions can be made for GFLs:
\vspace{-0.2em} 
\begin{itemize}
    \item \textbf{A1}: The dynamics and control mechanisms behind the DC boost converter, such as natural resource dynamics and wind pitch control, are disregarded. Thus, the power input $P^{F,In}$ in Fig. \ref{fig:GFLCont} remains constant, i.e., $P^{F,In} = P^{F,In0}$, where the subscript $\{\cdot\}^{0}$ indicates the steady-state value just prior to the disturbance.
    \item \textbf{A2}: The control setpoints $U^{Dc*}$, $Q^{F*}$, and $U^{F*}$ remain constant during post-disturbance dynamics.
    \item \textbf{A3}: The dynamics of high-bandwidth current control loops are neglected, implying that the actual and reference values of the converter current are equal, i.e., $i_{d}^{P*} = i_{d}^{P}$ and $i_{q}^{P*} = i_{q}^{P}$.
    \item \textbf{A4}: The q-axis control maintains terminal voltage and reactive power, so these values, as well as the q-axis current, remain constant after the disturbance, i.e., $Q^{F} = Q^{F0}$, $U^{F} = U^{F0}$, and $i_{q}^{P} = i_{q}^{P0}$.
\end{itemize}
\vspace{-0.2em} 

\begin{figure}[!t]
	\centering
	\includegraphics[width=0.46\textwidth]{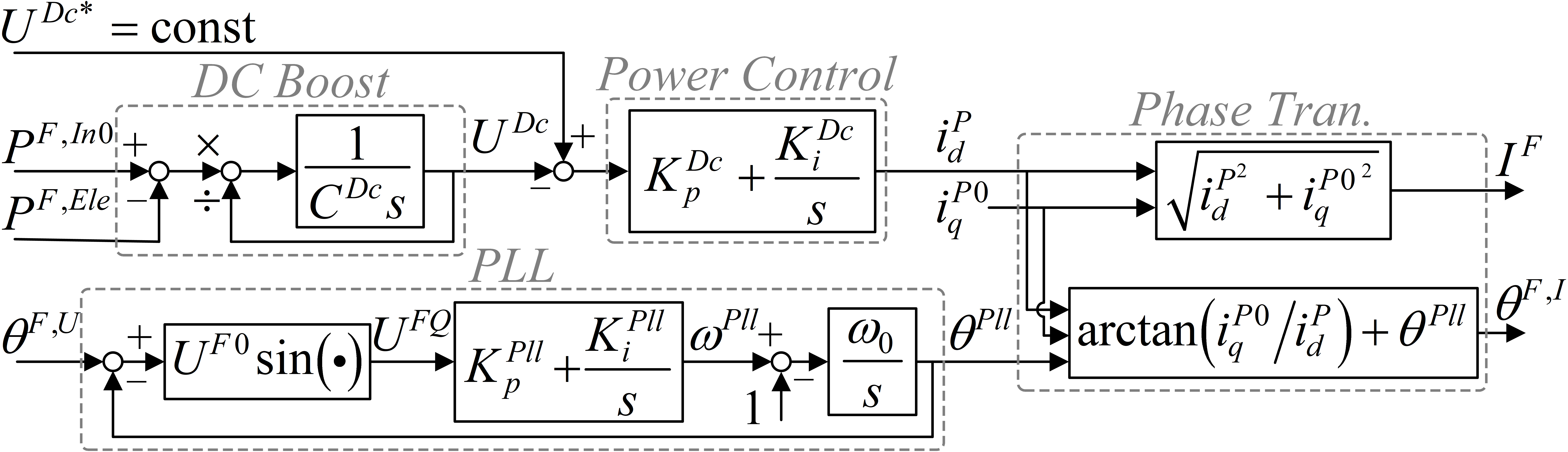}
    \vspace{-1.0em}
	\caption{GFL model in system frequency dynamics time scale.}
    \vspace{-1.8em}
	\label{fig:GFLSimp}
\end{figure}

Given the assumptions above, the detailed GFL model shown in Fig. \ref{fig:GFLCont} can be simplified to Fig. \ref{fig:GFLSimp}. Only the dynamics of the DC boost, d-axis power control, and PLL are retained.

As illustrated in Fig. \ref{fig:GFLSimp}, the state variables interfacing with the power grid include the d-axis current $i_{d}^{P}$ and the PLL phase angle $\theta^{Pll}$. Together, these form the amplitude $I^{F}$ and phase angle $\theta^{F,I}$ of the output current, which can be defined as the interface state variables of GFLs for simplicity:
\vspace{-0.5em} 
\begin{equation}\label{eq:DefSSF}
    \text{interface state variable: } I^{F} \angle \theta^{F,I}.
    \vspace{-0.7em} 
\end{equation}

\vspace{-1.4em} 
\subsection{On the GFL Dynamics Side\label{SubSec:GenePart}}
\vspace{-0.3em} 

As given in (\ref{eq:DefSSF}), the dynamics can be divided into two parts: the dynamics of the phase angle $\theta^{F,U}$ and the amplitude $I^{F}$.

\subsubsection{Dynamics of Angle} The model should be derived in a form similar to the dynamics of the phase angle of SGs, as described in the swing equation (\ref{eq:GeneSide}).

The input-output mappings of the DC boost, PLL, and phase transform modules in Fig. \ref{fig:GFLSimp} are nonlinear, and the inputs (including $U^{Dc*}$, $P^{F,In0}$, and $\theta^{F,U}$) driving the dynamics of $\theta^{F,U}$ are not the active power $P^{F,Ele}$. This complexity makes it challenging to represent Fig. \ref{fig:GFLSimp} using a simple swing equation. Therefore, the first step is to linearize Fig. \ref{fig:GFLSimp} and standardize the overall input. The process is outlined below:
\vspace{-0.2em} 
\begin{itemize}
    \item The DC boost and d-axis power control modules can be linearized as follows:
    \begin{subequations}\label{eq:LinDCp}
    \vspace{-0.5em} 
    \begin{equation}\label{eq:LinDCp_DC}
        \Delta P^{F,Ele} = U^{Dc0} C^{Dc} s \Delta U^{Dc},
        \vspace{-0.9em} 
    \end{equation}
    \vspace{-0.5em} 
    \begin{equation}\label{eq:LinDCp_APC}
        \Delta i_{d}^{P*} = \left( K_{p}^{Dc} + \frac{K_{i}^{Dc}}{s} \right) \Delta U^{Dc},
        \vspace{-0.7em} 
    \end{equation}
    \end{subequations}
    where the command $U^{Dc*}$ and wind power $P^{F,In0}$ are omitted, as they are constant according to the assumptions. These equations can then be aggregated as:
    \vspace{-0.5em} 
    \begin{equation}\label{eq:LinDC}
        \Delta i_{d}^{P} = \frac{K_{p}^{Dc} s + K_{i}^{Dc}}{U^{Dc0} C^{Dc} s^2} \Delta P^{F,Ele}.
        \vspace{-0.2em} 
    \end{equation}

    \item The dynamics of the PLL can be linearized as follows:
    \begin{subequations}\label{eq:LinPLLp}
    \vspace{-0.5em} 
    \begin{equation}\label{eq:LinPLLp_a}
        \Delta U^{FQ} = c^{Pll} \left(\Delta \theta^{F,U}-\Delta \theta^{Pll}\right),
        \vspace{-0.7em} 
    \end{equation}
    \vspace{-0.5em} 
    \begin{equation}\label{eq:LinPLLp_b}
        \Delta \omega^{Pll} = \left(K_{p}^{Pll} + \frac{K_{i}^{Pll}}{s}\right) \Delta U^{FQ},
        \vspace{-0.7em} 
    \end{equation}
    \vspace{-0.0em} 
    \begin{equation}\label{eq:LinPLLp_c}
        \Delta \theta^{Pll} = \frac{\omega^{0}}{s} \Delta \omega^{Pll},
        \vspace{-0.5em} 
    \end{equation}
    \end{subequations}
    where the linearization coefficient $c^{Pll}$ is provided in the Appendix \cite{AddDoc}. Then, (\ref{eq:LinPLLp}) can then be combined as:
    \vspace{-0.5em} 
    \begin{equation}\label{eq:LinPLL}
        \Delta \theta^{Pll} = \frac{c^{Pll} K_{p}^{Pll} s + c^{Pll} K_{i}^{Pll}}{s^2 + c^{Pll} K_{p}^{Pll} s + c^{Pll} K_{i}^{Pll}} \Delta \theta^{F,U},
        \vspace{-0.5em} 
    \end{equation}

    \item Similarly, the calculation logic of the angle phase transform can be linearized as follows:
    \vspace{-0.5em} 
    \begin{equation}\label{eq:LinPha}
        \Delta \theta^{F,I} = c^{Pi} \Delta i_{d}^{P} + \Delta \theta^{Pll},
        \vspace{-0.7em} 
    \end{equation}
    where the coefficient $c^{Pi}$ is provided in Appendix \cite{AddDoc}.
\end{itemize}
\vspace{-0.2em} 

Fig. \ref{fig:GFLSimp} is now linearized as shown in (\ref{eq:LinDC}), (\ref{eq:LinPLL}), and (\ref{eq:LinPha}). However, an undesired input, $\theta^{F,U}$, still exists in (\ref{eq:LinPLL}). To resolve this, $\theta^{F,U}$ needs to be expressed as a function of the power $P^{F,Ele}$. Since the current $I^{F} \angle \theta^{F,I}$ is the overall output and can be assumed to be known based on (\ref{eq:LinDC}), (\ref{eq:LinPLL}), and (\ref{eq:LinPha}), the relationship between the current and terminal voltage can be used to map the relation between $\theta^{F,U}$ and $P^{F,Ele}$:
\vspace{-0.5em} 
\begin{equation}\label{eq:Power}
    P^{F,Ele} = U^{F0} I^{F} \cos\left( \theta^{F,U} - \theta^{F,I} \right).
    \vspace{-0.7em} 
\end{equation}
Given $I^{F} = \sqrt{{i_{d}^{P}}^2 + {i_{q}^{P0}}^2}$, (\ref{eq:Power}) can be linearized as follows:
\vspace{-0.3em} 
\begin{equation}\label{eq:LinPow}
    \Delta P^{F,Ele} = c^{Ei} \Delta i_{d}^{P} + c^{Ep} \Delta \theta^{F,U} - c^{Ep} \Delta \theta^{F,I} ,
    \vspace{-0.5em} 
\end{equation}
where coefficients $c^{Ei}$ and $c^{Ep}$ are provided in Appendix \cite{AddDoc}.

Now, (\ref{eq:LinDC}), (\ref{eq:LinPLL}), (\ref{eq:LinPha}), and (\ref{eq:LinPow}) together form the linearized dynamics of $\theta^{F,I}$ with an overall input of $P^{F,Ele}$. The dynamics of $\theta^{F,I}$ can be combined into transfer functions as shown in (\ref{eq:PhCombAll}), where the structure of (\ref{eq:LinPha}) is preserved and the dynamics of the two state variables $i_{d}^{P}$ and $\theta^{Pll}$ are separated, as follows:
\vspace{-0.9em} 
\begin{equation}\label{eq:PhCombAll}
    \theta^{F,I} = c^{Pi} \underbrace{J^{Id}\left(s\right) \Delta P^{F,Ele}}_{\mathrm{dynamics \ of \ } i_{d}^{P}} + \underbrace{J^{Pll}\left(s\right) \Delta P^{F,Ele}}_{\mathrm{dynamics \ of \ } \theta^{Pll}},
    \vspace{-0.5em} 
\end{equation}
where $J^{Id}\left(s\right)$ and $J^{Pll}\left(s\right)$ are detailed in Appendix \cite{AddDoc}.

A next step is to introduce an equivalent frequency for (\ref{eq:PhCombAll}), similar to the rotor frequency in (\ref{eq:GeneSide}). Given the two dynamic components in (\ref{eq:PhCombAll}), two parts of the equivalent frequency are separated, as follows:
\begin{subequations}\label{eq:SSDynW}
\vspace{-0.5em} 
\begin{equation}\label{eq:SSDynW_ph}
    \vspace{-0.5em} 
    \frac{\mathrm{d} \Delta \theta^{F,I}}{\mathrm{d} t} = \omega^{0} \Delta \omega^{F},
    \vspace{-0.7em} 
\end{equation}
\vspace{-0.0em} 
\begin{equation}\label{eq:SSDynW_w}
    \Delta \omega^{F} = c^{Pi} \Delta \omega^{F,Id} + \Delta \omega^{F,Pll},
    \vspace{-0.7em} 
\end{equation}
\end{subequations}
where $\omega^{F}$ is the overall equivalent frequency, and $\omega^{F,Id}$ and $\omega^{F,Pll}$ are the components corresponding to $i_{d}^{P}$ and $\theta^{Pll}$.

Comparing (\ref{eq:GeneSide}) and (\ref{eq:SSDynW}), (\ref{eq:SSDynW}) can then be transformed into (\ref{eq:SSDynP}), where the superscript $\{\cdot\}$ can be $Id$ or $Pll$, corresponding to the components $i_{d}^{P}$ and $\theta^{Pll}$. Two types of frequency components are included, referred to as a continuous component $\omega_{c}^{F}$ and a discontinuous component $\omega_{d}^{F}$:
\begin{subequations}\label{eq:SSDynP}
\vspace{-0.5em} 
\begin{equation}\label{eq:SSDynP_w}
    \Delta \omega^{F,\{\cdot\}} = \Delta \omega_{c}^{F,\{\cdot\}} + \Delta \omega_{d}^{F,\{\cdot\}},
    \vspace{-0.7em} 
\end{equation}
\vspace{-0.9em} 
\begin{equation}\label{eq:SSDynP_H}
    2 H^{F,\{\cdot\}} \frac{\mathrm{d} \Delta \omega_{c}^{F,\{\cdot\}}}{\mathrm{d} t} = \Delta P^{F,Mec,\{\cdot\}} - \Delta P^{F,Ele},
    \vspace{-0.7em} 
\end{equation}
\vspace{-0.5em} 
\begin{equation}\label{eq:SSDynP_Pm}
    \Delta P^{F,Mec,\{\cdot\}} = J^{F,\{\cdot\}}\left(s\right) \Delta \omega_{c}^{F,\{\cdot\}},
    \vspace{-0.7em} 
\end{equation}
\vspace{-0.9em} 
\begin{equation}\label{eq:SSDynP_dw}
    \Delta \omega_{d}^{F,\{\cdot\}} = L^{F,\{\cdot\}} \Delta P^{F,Ele},
    \vspace{-0.7em} 
\end{equation}
\end{subequations}
where $H^{F}$ denotes the equivalent inertia; $J^{F}\left(s\right)$ represents the transfer function of the equivalent governor; $\Delta P^{F,Mec}$ is the equivalent mechanical power; the coefficient $L^{F}$, defined as the proportionality coefficient, arises because the denominator is one degree higher than the numerator of $J^{Id}\left(s\right)$ and $J^{Pll}\left(s\right)$ in (\ref{eq:PhCombAll}). The expressions for above are given in Appendix \cite{AddDoc}.

A summary of the above dynamics is illustrated in Figure 1 in the Appendix \cite{AddDoc}.

\subsubsection{Dynamics of Amplitude} Since the amplitude dynamics play a minor role in frequency dynamics analysis, it can be expressed simply as in Appendix \cite{AddDoc}.

\vspace{-1.2em} 
\subsection{On the Network Side\label{SubSec:NetPart}}
\vspace{-0.3em} 

Similar to (\ref{eq:NetwSide}), the active power $P^{F,Ele}$ of GFLs should be expressed in terms of the interface state variables $E^{G} \angle \delta^{G}$ for SGs and $I^{F} \angle \theta^{F,I}$ for GFLs. The same formulation applies to the active power $P^{G,Ele}$ of SGs.

The power equation (\ref{eq:NetwSide}) is derived from the system's power flow relationships. Therefore, when incorporating GFLs, the initial step should involve establishing the power flow relationships. Let $E^{G} \angle \delta^{G}$, $I^{F} \angle \theta^{F,I}$, and $U^{F} \angle \theta^{F,U}$ represent the phasors, denoted as $\dot{E}^{G}$, $\dot{I}^{F}$, and $\dot{U}^{F}$, respectively. These phasors can be organized as vectors: $\boldsymbol{\dot{E}}^{G} \in \mathbb{C}_{N^{G} \times 1}$, $\boldsymbol{\dot{I}}^{F} \in \mathbb{C}_{N^{F} \times 1}$, and $\boldsymbol{\dot{U}}^{F} \in \mathbb{C}_{N^{F} \times 1}$, where $N^{F}$ is the total number of GFLs. The power flow described by the system admittance matrix can then be expressed as follows:
\vspace{-0.5em} 
\begin{equation}\label{eq:IYloadU}
    \left[\begin{matrix}
        \boldsymbol{\dot{I}}^{G} \\
        \boldsymbol{\dot{I}}^{F}
    \end{matrix}\right] = 
    \left[\begin{matrix}
        \boldsymbol{\dot{\tilde{Y}}}^{GG} & \boldsymbol{\dot{\tilde{Y}}}^{GF} \\
        \boldsymbol{\dot{\tilde{Y}}}^{FG} & \boldsymbol{\dot{\tilde{Y}}}^{FF}
    \end{matrix}\right] 
    \left[\begin{matrix}
        \boldsymbol{\dot{E}}^{G} \\
        \boldsymbol{\dot{U}}^{F}
    \end{matrix}\right],
    \vspace{-0.7em} 
\end{equation}
where $\boldsymbol{\dot{I}}^{G}$ is current vector of SGs; $\boldsymbol{\dot{\tilde{Y}}}$ is the node-eliminated admittance matrix. This means the load connected to the network node is given by its admittance or impedance and is included in $\boldsymbol{\dot{\tilde{Y}}}$. Derivations are provided in Appendix \cite{AddDoc}.

In (\ref{eq:IYloadU}), the state variables $\dot{E}^{G}$ of SGs and $\dot{I}^{F}$ of GFLs are positioned on opposite sides of the equation. This complicates the direct correlation between these state variables. To address this, $\dot{I}^{F}$ is moved to the right side of (\ref{eq:IYloadU}), resulting in:
\vspace{-0.5em} 
\begin{equation}\label{eq:IYUibr}
    \left[\begin{matrix}
        \boldsymbol{\dot{I}}^{G} \\
        \boldsymbol{\dot{U}}^{F}
    \end{matrix}\right] = 
    \left[\begin{matrix}
        \boldsymbol{\dot{Y}}^{eq} & \boldsymbol{\dot{T}}^{eq,I} \\
        \boldsymbol{\dot{T}}^{eq,U} & \boldsymbol{\dot{Z}}^{eq}
    \end{matrix}\right] 
    \left[\begin{matrix}
        \boldsymbol{\dot{E}}^{G} \\
        \boldsymbol{\dot{I}}^{F}
    \end{matrix}\right],
    \vspace{-0.7em} 
\end{equation}
where the $\boldsymbol{\dot{Y}}^{eq}$, $\boldsymbol{\dot{T}}^{eq,I}$, $\boldsymbol{\dot{T}}^{eq,U}$, and $\boldsymbol{\dot{Z}}^{eq}$ form the new matrix connecting current and voltage. The symbols of the above connection matrices are aligned with their physical meanings: $\dot{Y}^{eq}$ represents the equivalent admittance, $\dot{Z}^{eq}$ denotes the equivalent impedance, and $\dot{T}^{eq}$ is defined as the transfer parameter connecting currents or voltages themselves. These coefficients are expressed as:
\begin{subequations}\label{eq:IYUibrCoefs}
\vspace{-0.5em} 
\begin{equation}\label{eq:IYUibrCoefs_Y}
    \boldsymbol{\dot{Y}}^{eq} = \boldsymbol{\dot{\tilde{Y}}}^{GG} - {\boldsymbol{\dot{\tilde{Y}}}^{FF}}^{-1} \boldsymbol{\dot{\tilde{Y}}}^{GF} \boldsymbol{\dot{\tilde{Y}}}^{FG},
    \vspace{-0.7em} 
\end{equation}
\vspace{-0.9em} 
\begin{equation}\label{eq:IYUibrCoefs_T1}
    \boldsymbol{\dot{T}}^{eq,I} = {\boldsymbol{\dot{\tilde{Y}}}^{FF}}^{-1} \boldsymbol{\dot{\tilde{Y}}}^{GF},
    \vspace{-0.7em} 
\end{equation}
\vspace{-0.9em} 
\begin{equation}\label{eq:IYUibrCoefs_T2}
    \boldsymbol{\dot{T}}^{eq,U} = -{\boldsymbol{\dot{\tilde{Y}}}^{FF}}^{-1}\boldsymbol{\dot{\tilde{Y}}}^{FG},
    \vspace{-0.7em} 
\end{equation}
\vspace{-0.9em} 
\begin{equation}\label{eq:IYUibrCoefs_Z}
    \boldsymbol{\dot{Z}}^{eq} = {\boldsymbol{\dot{\tilde{Y}}}^{FF}}^{-1}.
    \vspace{-0.9em} 
\end{equation}
\end{subequations}
Given that the admittance matrix is symmetric, $\boldsymbol{\dot{\tilde{Y}}}^{GF}$ and $\boldsymbol{\dot{\tilde{Y}}}^{FG}$ in (\ref{eq:IYloadU}) are also symmetric, with the relationship denoted as $\boldsymbol{\dot{\tilde{Y}}}^{GF} = \Phi \left( \boldsymbol{\dot{\tilde{Y}}}^{FG} \right) $. According to (\ref{eq:IYUibrCoefs_T1}) and (\ref{eq:IYUibrCoefs_T2}), the symmetry holds: $\boldsymbol{\dot{T}}^{eq,I} = -\Phi \left( \boldsymbol{\dot{T}}^{eq,U} \right)$ holds. Letting $\boldsymbol{\dot{T}}^{eq,I}$ be expressed as $\boldsymbol{\dot{T}}^{eq}$, (\ref{eq:IYUibrCoefs}) can then be rewritten as follows:
\vspace{-0.5em} 
\begin{equation}\label{eq:IYUsym}
    \left[\begin{matrix}
        \boldsymbol{\dot{I}}^{G} \\
        \boldsymbol{\dot{U}}^{F}
    \end{matrix}\right] = 
    \left[\begin{matrix}
        \boldsymbol{\dot{Y}}^{eq} & \boldsymbol{\dot{T}}^{eq} \\
        -\Phi \left(\boldsymbol{\dot{T}}^{eq}\right) & \boldsymbol{\dot{Z}}^{eq}
    \end{matrix}\right] 
    \left[\begin{matrix}
        \boldsymbol{\dot{E}}^{G} \\
        \boldsymbol{\dot{I}}^{F}
    \end{matrix}\right].
    \vspace{-0.7em} 
\end{equation}

The power flow equations connecting the interface state variables can be derived based on (\ref{eq:IYUsym}) considering (i) the active power of a single generator $P = \mathrm{Re} \left(\dot{U} \times \mathrm{conjugate}\left(\dot{I}\right)\right)$ and (ii) let the elements $\dot{Y}^{eq} = G^{eq} + j B^{eq}$, $\dot{T}^{eq} = V^{eq} + j W^{eq}$, and $\dot{Z}^{eq} = R^{eq} + j X^{eq}$.

First, the active power of SGs, $P^{G,Ele}$, are given in (\ref{eq:PexprG}), consisting of (i) the local power $P^{G,Loc}$, (ii) the power transferred through an equivalent tie line with other SGs, $P^{G,Tie,G}$, which connects $E^{G} \angle \delta^{G}$ of SGs, and (iii) the power exchanged with other GFLs, $P^{G,Tie,F}$, connecting the local $E^{G} \angle \delta^{G}$ with $I^{F} \angle \theta^{F,I}$ of GFLs:
\begin{subequations}\label{eq:PexprG}
\vspace{-0.5em} 
\begin{equation}\label{eq:PexprG_E}
    P_{i^{G}}^{G,Ele} = P_{i^{G}}^{G,Loc} + P_{i^{G}}^{G,Tie,G} + P_{i^{G}}^{G,Tie,F},
    \vspace{-1.0em} 
\end{equation}
\vspace{-0.5em} 
\begin{equation}\label{eq:PexprG_L}
    P_{i^{G}}^{G,Loc} = {E_{i^{G}}^{G}}^2 G_{i^{G},i^{G}}^{eq},
    \vspace{-1.0em} 
\end{equation}
\vspace{-0.5em} 
\begin{equation}\label{eq:PexprG_TG}
\begin{aligned}
    P_{i^{G}}^{G,Tie,G} = \! \! \! \sum_{\substack{\forall j^{G} \in N^{G} \\ j^{G} \neq i^{G}}} \! \! E_{i^{G}}^{G} E_{j^{G}}^{G} & \left( G_{i^{G},j^{G}}^{eq} \cos \left(\delta_{i^{G}}^{G} - \delta_{j^{G}}^{G} \right) \right. \\
    \\[-3.3em]
    & \left. + B_{i^{G},j^{G}}^{eq} \sin \left(\delta_{i^{G}}^{G} - \delta_{j^{G}}^{G} \right)\right), \! \! \! \! \!
\end{aligned}
\vspace{-0.2em} 
\end{equation}
\vspace{-0.5em} 
\begin{equation}\label{eq:PexprG_TF}
\begin{aligned}
    P_{i^{G}}^{G,Tie,F} = \sum_{\forall i^{F} \in N^{F}} \! \! E_{i^{G}}^{G} I_{i^{F}}^{F} & \left( - V_{i^{G},i^{F}}^{eq} \cos \left(\delta_{i^{G}}^{G} - \theta_{i^{F}}^{F,I} \right) \right. \! \! \! \! \! \! \! \\
    \\[-2.3em]
    & \left. + W_{i^{G},i^{F}}^{eq} \sin \left(\delta_{i^{G}}^{G} - \theta_{i^{F}}^{F,I} \right)\right), \! \! \! \! \! \! \!
\end{aligned}
\vspace{-0.7em} 
\end{equation}
\end{subequations}
where $i^{F}$ is the index for GFLs.

Similarly, the active power $P^{F,Ele}$ of GFLs can be derived. It also has three components: (i) the local power $P^{F,Loc}$, (ii) the power transferred through an equivalent tie line with other GFLs, $P^{F,Tie,F}$ connecting $I^{F} \angle \theta^{F,I}$, and (iii) the power exchanged with other SGs, $P^{F,Tie,G}$, which connects $E^{G} \angle \delta^{G}$ of SGs with $I^{F} \angle \theta^{F,I}$, as follows:
\begin{subequations}\label{eq:PexprF}
\vspace{-0.5em} 
\begin{equation}\label{eq:PexprF_E}
    P_{i^{F}}^{F,Ele} = P_{i^{F}}^{F,Loc} + P_{i^{F}}^{F,Tie,F} + P_{i^{F}}^{F,Tie,G},
    \vspace{-1.0em} 
\end{equation}
\vspace{-0.5em} 
\begin{equation}\label{eq:PexprF_L}
    P_{i^{F}}^{F,Loc} = {I_{i^{F}}^{F}}^2 R_{i^{F},i^{F}}^{eq},
    \vspace{-0.9em} 
\end{equation}
\vspace{-0.5em} 
\begin{equation}\label{eq:PexprF_TF}
\begin{aligned}
    P_{i^{F}}^{F,Tie,F} = \sum_{\substack{\forall j^{F} \in N^{F} \\ j^{F} \neq i^{F}}} I_{i^{F}}^{F} I_{j^{F}}^{F} & \left(R_{i^{F},j^{F}}^{eq} \cos \left(\theta_{i^{F}}^{F,I} - \theta_{i^{F}}^{F,I}\right) \right. \! \! \! \! \! \! \\
    \\[-3.3em]
    & \left. + X_{i^{F},j^{F}}^{eq} \sin \left(\theta_{i^{F}}^{F} - \theta_{j^{F}}^{F} \right)\right), \! \! \! \! \! \!
\end{aligned}
\vspace{-0.1em} 
\end{equation}
\vspace{-0.5em} 
\begin{equation}\label{eq:PexprF_TG}
\begin{aligned}
    P_{i^{F}}^{F,Tie,G} = \sum_{\forall i^{G} \in N^{G}} I_{i^{F}}^{F} E_{i^{G}}^{G} & \left(- V_{i^{F},i^{G}}^{eq} \cos \left(\theta_{i^{F}}^{F,I} - \delta_{i^{G}}^{G} \right) \right. \! \! \! \! \! \! \\
    \\[-2.3em]
    & \left. + W_{i^{F},i^{G}}^{eq} \sin \left( \theta_{i^{F}}^{F,I} - \delta_{i^{G}}^{G} \right) \! \right) \!, \! \! \! \! \! \! \!
\end{aligned}
\vspace{-0.5em} 
\end{equation}
\end{subequations}
where $j^{F}$ is the index for GFLs.

\vspace{-1.5em} 
\subsection{A Summary\label{SubSec:Sum}}
\vspace{-0.3em} 

\begin{figure}[!t]
	\centering
	\includegraphics[width=0.46\textwidth]{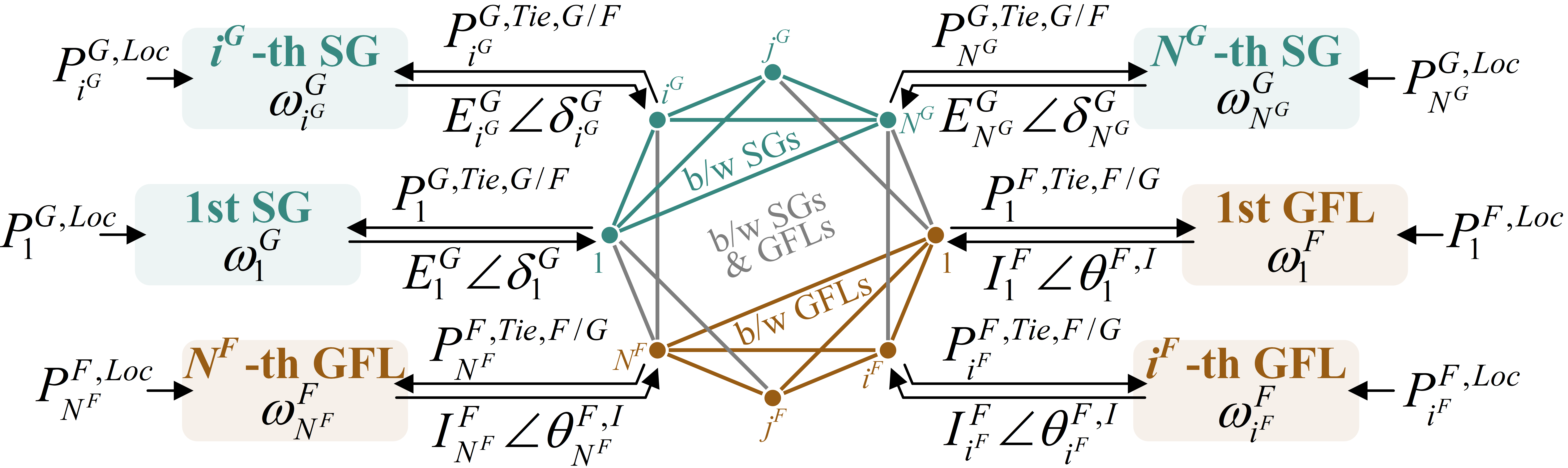}
    \vspace{-1.0em}
	\caption{Multi-generator dynamics in power systems with GFLs.}
    \vspace{-1.8em}
	\label{fig:mmCOIGFLs}
\end{figure}

A summary of the above is illustrated in Fig. \ref{fig:mmCOIGFLs}. On the generator dynamics side, the dynamics of $E^{G} \angle \delta^{G}$ under the active power $P^{G,Ele}$ for SGs follow the same form as in a traditional system, as given in (\ref{eq:GeneSide}). The dynamics of the GFL state variable $I^{F} \angle \theta^{F,I}$ under the active power $P^{F,Ele}$ are introduced, as given in (\ref{eq:SSDynW})-(\ref{eq:SSDynP}). On the network side, active power is transferred through virtual tie lines that connect $E^{G} \angle \delta^{G}$ or $I^{F} \angle \theta^{F,I}$, as described in (\ref{eq:PexprG}) and (\ref{eq:PexprF}). These connections include: (i) $P^{G,Tie,G}$ between SGs themselves, marked as ``b/w SGs''; (ii) $P^{F,Tie,F}$ between GFLs themselves, marked as ``b/w GFLs''; and (iii) $P^{G,Tie,F}$ and $P^{F,Tie,G}$ between SGs and GFLs, marked as ``b/w SGs \& GFLs''.

\vspace{-1.3em} 
\section{Effect of GFLs on COI Frequency Dynamics\label{Sec:GFLeffe}}
\vspace{-0.1em} 

The COI frequency dynamics is aggregated in Section \ref{SubSec:SGvsGFL} based on the system multi-generator model. The impact of GFLs on COI dynamics is discussed in detail in Section \ref{SubSec:GFLimpc}.

\vspace{-1.1em} 
\subsection{COI Frequency Dynamics Considering GFLs\label{SubSec:SGvsGFL}}
\vspace{-0.3em} 

As discussed in Section \ref{Sec:ProbForm}, forming the COI dynamics requires (i) inertia-like dynamics on the generator side and (ii) symmetric tying powers on the network side. In Fig. \ref{fig:mmCOIGFLs}, SGs follow the inertia motion law, and GFLs are modeled to follow a similar law. The symmetry of the tying powers is analyzed below for three components:
\vspace{-0.2em} 
\begin{itemize}
    \item \textbf{Tying between two SGs in (\ref{eq:PexprG_TG})}: The terms $G^{eq}$ and $B^{eq}$ in $\boldsymbol{\dot{Y}}^{eq}$ (\ref{eq:IYUsym}) determine the tying power. The $G^{eq}$ terms are relatively small compared to $B^{eq}$ due to the dominance of reactance over resistance in power systems. As in traditional systems, the symmetric property $B_{i^{G},j^{G}}^{eq} = B_{j^{G},i^{G}}^{eq}$ ensures that the tying powers between any pair of SGs have equal magnitudes but opposite directions:
    \vspace{-0.5em} 
    \begin{equation}\label{eq:TwoSGs}
    \begin{aligned}
        & \underbrace{E_{i^{G}}^{G} E_{j^{G}}^{G} B_{i^{G},j^{G}}^{eq} \sin \left(\delta_{i^{G}}^{G} - \delta_{j^{G}}^{G} \right)}_{\mathrm{at} \ i^{G}\mathrm{-th} \ \mathrm{SG} \ \mathrm{side}} \\
        & + \underbrace{E_{j^{G}}^{G} E_{i^{G}}^{G} B_{j^{G},i^{G}}^{eq} \sin \left(\delta_{j^{G}}^{G} - \delta_{i^{G}}^{G} \right)}_{\mathrm{at} \ j^{G}\mathrm{-th} \ \mathrm{SG} \ \mathrm{side}} = 0.
    \end{aligned}
    \vspace{-0.5em} 
    \end{equation}
    
    \item \textbf{Tying between SGs and GFLs in (\ref{eq:PexprG_TF}) and (\ref{eq:PexprF_TG})}: The $\boldsymbol{\dot{T}}^{eq}$ matrix in (\ref{eq:IYUsym}) requires attention. Since $\boldsymbol{\dot{T}}^{eq}$ is formed by even times of products of admittance matrices, as shown in (\ref{eq:IYUibrCoefs_T1}) and (\ref{eq:IYUibrCoefs_T2}), the magnitudes of its real and imaginary parts are inversely compared to those of the original admittance matrices. Note that the $V^{eq}$ terms are larger than the $W^{eq}$ terms. The symmetry is preserved, as $V_{i^{G},i^{F}}^{eq} = V_{i^{F},i^{G}}^{eq}$. Repeating the above calculations:
    \vspace{-0.5em} 
    \begin{equation}\label{eq:SGsGFLs}
    \begin{aligned}
        & - \! \underbrace{ E_{i^{G}}^{G} I_{i^{F}}^{F} V_{i^{G},i^{F}}^{eq} \! \cos \! \left( \delta_{i^{G}}^{G} \! - \! \theta_{i^{F}}^{F,I} \right) }_{\mathrm{at} \ i^{G}\mathrm{-th} \ \mathrm{SG} \ \mathrm{side}} - \underbrace{I_{i^{F}}^{F} E_{i^{G}}^{G} V_{i^{F},i^{G}}^{eq} \times}_{\mathrm{at} \ i^{F}\mathrm{-th} \ \mathrm{GFL}} \! \! \! \! \! \! \! \\
        & \underbrace{\cos \! \left( \theta_{i^{F}}^{F,I} \! -  \! \delta_{i^{G}}^{G} \right)}_{\mathrm{side}} \! = - 2 E_{i^{G}}^{G} I_{i^{F}}^{F} V_{i^{G},i^{F}}^{eq} \! \cos \! \left( \delta_{i^{G}}^{G} \! - \! \theta_{i^{F}}^{F,I} \right) \!. \! \! \! \! \! \! \!
    \end{aligned}
    \vspace{-0.4em} 
    \end{equation}
    Unlike in (\ref{eq:TwoSGs}), the tying powers between the SG and GFL have equal magnitudes but in the same direction, preventing them from being aggregated into a single COI.

    \item \textbf{Tying between two GFLs in (\ref{eq:PexprF_TF})}: Since the interaction between SGs and GFLs does not follow the COI pattern, GFLs must be treated separately, regardless of the form of the tying powers between two GFLs. Thus, their dynamics cannot be directly aggregated with SGs.
\end{itemize}
\vspace{-0.2em} 

Based on the above, the model of COI frequency dynamics can be established, where all SGs are aggregated into the SG COI frame, while connections between COI and all GFLs are preserved. First, the state variable of the SG COI interfacing with GFLs can be represented as the average of EMFs across all SGs, as follows:
\vspace{-0.5em} 
\begin{equation}\label{eq:COISS}
    \text{interface state variable: } E^{Coi} \angle \delta^{Coi}.
    \vspace{-0.7em} 
\end{equation}

The SG COI model includes both the generator dynamics of SG COI and the power network linking the COI and GFLs. On the generator side, the dynamics are given in (\ref{eq:PropCOI_H}) and (\ref{eq:PropCOI_d}), where the COI angle $\delta^{Coi}$ is driven by the COI active power $P^{Coi,Ele}$. The amplitude part, $E^{Coi}$, remains constant. On the network side, $P^{Coi,Ele}$ is composed of (i) the local power $P^{Coi,Loc}$ given in (\ref{eq:PropCOI_L}) and (ii) the tying power with GFLs, linked by the COI state variables $E^{Coi} \angle \delta^{Coi}$ and $I^{F} \angle \theta^{F,I}$ of GFLs, as given in (\ref{eq:PropCOI_P}). The above equations are as follows:
\begin{subequations}\label{eq:PropCoi}
\vspace{-0.7em} 
\begin{equation}\label{eq:PropCOI_H}
    2 H^{Coi} \frac{\mathrm{d} \omega^{Coi}}{\mathrm{d} t} = P^{Coi,Mec} - P^{Coi,Ele},
    \vspace{-0.4em} 
\end{equation}
\vspace{-0.5em} 
\begin{equation}\label{eq:PropCOI_d}
    \frac{\mathrm{d} \delta^{Coi}}{\mathrm{d} t} = \omega^{0} \left( \omega^{Coi} - 1 \right),
    \vspace{-0.7em} 
\end{equation}
\vspace{-0.5em} 
\begin{equation}\label{eq:PropCOI_E}
    P^{Coi,Ele} = P^{Coi,Loc} + P^{Coi,Tie,F},
    \vspace{-1.0em} 
\end{equation}
\vspace{-0.5em} 
\begin{equation}\label{eq:PropCOI_L}
    P^{Coi,Loc} = {E^{Coi}}^2 G^{eq\prime},
    \vspace{-0.7em} 
\end{equation}
\vspace{-0.8em} 
\begin{equation}\label{eq:PropCOI_P}
\begin{aligned}
    P^{Coi,Tie,F} \! \! \! = \! \! \! \! \sum_{\forall i^{F} \in N^{F}} \! \! \! \! E^{Coi} I_{i^{F}}^{F} & \left( - V_{i^{F}}^{eq\prime} \cos \left(\delta^{Coi} - \theta_{i^{F}}^{F,I} \right) \right. \! \! \! \! \! \! \! \! \! \! \! \\
    \\[-2.3em]
    & \left. + W_{i^{F}}^{eq\prime} \sin \left(\delta^{Coi} - \theta_{i^{F}}^{F,I} \right) \! \right) \!, \! \! \! \! \! \! \! \! \! \! \!
\end{aligned}
\vspace{-0.4em} 
\end{equation}
\end{subequations}
where the calculations for $H^{Coi}$ and $\omega^{Coi}$ are given in (\ref{eq:TradCOIfH}); the terms $G^{eq\prime}$, $V^{eq\prime}$, and $W^{eq\prime}$ retain the same physical meanings as in (\ref{eq:PexprG}), with the distinction that one end of the connection is now the SG COI instead of a single SG. Detailed calculations of $G^{eq\prime}$, $V^{eq\prime}$, and $W^{eq\prime}$ are provided in the Appendix \cite{AddDoc}.

The generator dynamics of GFLs still follows Section \ref{SubSec:GenePart}. The active power of GFLs, $P^{F,Ele}$, consists of (i) the local power $P^{F,Loc}$ given in (\ref{eq:PGFLs_L}), (ii) the tying power with other GFLs, $P_{i^{F}}^{F,Tie,F}$, linked by $I^{F} \angle \theta^{F,I}$ given in (\ref{eq:PGFLs_TF}), and (iii) the tying power with the SG COI, linked by $I^{F} \angle \theta^{F,I}$ and $E^{Coi} \angle \delta^{Coi}$ given in (\ref{eq:PGFLs_T}), as follows:
\begin{subequations}\label{eq:PGFLs}
\vspace{-0.5em} 
\begin{equation}\label{eq:PGFLs_E}
    P_{i^{F}}^{F,Ele} = P_{i^{F}}^{F,Loc} + P_{i^{F}}^{F,Tie,F} + P_{i^{F}}^{F,Tie,Coi},
    \vspace{-0.7em} 
\end{equation}
\vspace{-0.5em} 
\begin{equation}\label{eq:PGFLs_L}
    P_{i^{F}}^{F,Loc} = {I_{i^{F}}^{F}}^2 R_{i^{F},i^{F}}^{eq\prime},
    \vspace{-0.7em} 
\end{equation}
\vspace{-0.5em} 
\begin{equation}\label{eq:PGFLs_TF}
\begin{aligned}
    P_{i^{F}}^{F,Tie,F} = \sum_{\substack{\forall j^{F} \in N^{F} \\ j^{F} \neq i^{F}}} I_{i^{F}}^{F} I_{j^{F}}^{F} & \left(R_{i^{F},j^{F}}^{eq\prime} \cos \left(\theta_{i^{F}}^{F,I} - \theta_{i^{F}}^{F,I}\right) \right. \! \! \! \! \! \! \\
    \\[-3.3em]
    & \left. + X_{i^{F},j^{F}}^{eq\prime} \sin \left(\theta_{i^{F}}^{F} - \theta_{j^{F}}^{F} \right) \! \right) \!, \! \! \! \! \! \!
\end{aligned}
\vspace{-0.0em} 
\end{equation}
\vspace{-0.3em} 
\begin{equation}\label{eq:PGFLs_T}
\begin{aligned}
    P_{i^{F}}^{F,Tie,Coi} = I_{i^{F}}^{F} E^{Coi} & \left(- V_{j^{F}}^{eq\prime} \cos \left(\theta_{i^{F}}^{F,I} - \delta^{Coi} \right) \right. \\
    & \left. + W_{i^{F}}^{eq\prime} \sin \left( \theta_{i^{F}}^{F,I} - \delta^{Coi} \right)\right),
\end{aligned}
\vspace{-0.4em} 
\end{equation}
\end{subequations}
where the terms $R^{eq\prime}$ and $X^{eq\prime}$, considering the COI introduction, retain the same physical meanings as in (\ref{eq:PexprG}), with the detailed calculations provided in the Appendix \cite{AddDoc}.

\begin{figure}[!t]
	\centering
	\includegraphics[width=0.46\textwidth]{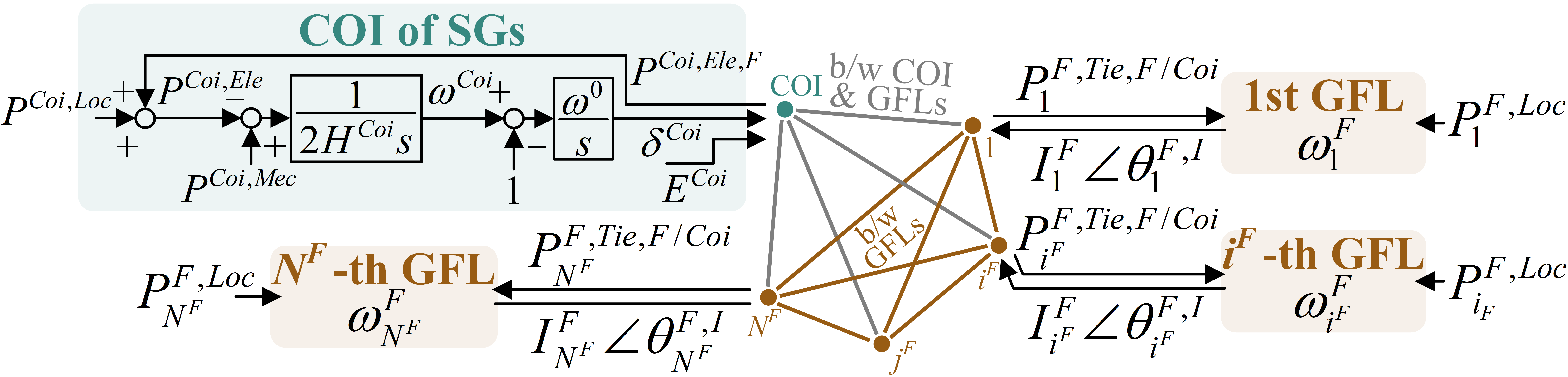}
    \vspace{-1.0em}
	\caption{COI frequency dynamics in power systems with GFLs.}
    \vspace{-1.8em}
	\label{fig:COIGFLs}
\end{figure}

In summary, the COI frequency dynamics with GFLs, as given in (\ref{eq:PropCoi}) and (\ref{eq:PGFLs}), are illustrated in Fig. \ref{fig:COIGFLs}. The equivalent ties between the COI and GFLs are labeled as ``b/w COI \& GFLs,'' and those between GFLs are labeled as ``b/w GFLs.''

\vspace{-0.9em} 
\subsection{Effect of GFLs on COI Frequecny Dynamics\label{SubSec:GFLimpc}}
\vspace{-0.1em} 

\subsubsection{How GFLs Affect COI Frequency Dynamics} Although the GFL has inertia $H^{F}$ and other dynamics affecting its equivalent frequency $\omega^{F}$, as shown Section \ref{SubSec:GenePart}, $\omega^{F}$ does not directly contribute to the COI frequency $\omega^{COI}$. Instead, it impacts $\omega^{COI}$ through the tying power $P^{Coi,Tie,F}$ between the state variable $I^{F} \angle \theta^{F,I}$ of the GFL and $E^{COI} \angle \delta^{COI}$ in (\ref{eq:PropCOI_P}). The mechanism by which $I^{F} \angle \theta^{F,I}$ affects $P^{Coi,Tie,F}$ can be discussed further:
\vspace{-0.2em} 
\begin{itemize}
    \item \textbf{Through angle $\boldsymbol{\theta^{F,I}}$}: $P^{Coi,Tie,F}$ is determined by the angle difference between $\theta^{F,I}$ and $\delta^{COI}$. The rotor-like dynamics of $\omega^{F}$ described in (\ref{eq:SSDynW})-(\ref{eq:SSDynP}) determines $\theta^{F,I}$, then interact with COI frequency $\omega^{COI}$.
    \item \textbf{Through amplitude $\boldsymbol{I^{F}}$}: $P^{Coi,Tie,F}$ is also affect by the amplitede $I^{F}$. However, the amplitude variation is less significant compared to the angle dynamics.
\end{itemize}
\vspace{-0.2em} 

The degree of GFL impact is determined by the transfer parameters $V_{eq}$ and $W_{eq}$. Due to the inverse of the admittance matrix in the product operation in (\ref{eq:IYUibrCoefs_T1}) or (\ref{eq:IYUibrCoefs_T2}), the values of $V_{eq}$ and $W_{eq}$ are low, indicating that the connection between the GFL and COI is weak. 

\subsubsection{GFL Local Dynamics} In (\ref{eq:SSDynW})-(\ref{eq:SSDynP}), the dynamics is determined by three components: the equivalent inertia $H^{F}$, the equivalent governor $J^{F}\left(s\right)$, and the so-called proportional coefficient $L^{F}$. In addition, two frequency components are included in (\ref{eq:SSDynW_w}): $\omega^{F,Id}$, which is determined by the dynamics of d-axis power control and DC boost, and $\omega^{F,Pll}$, which is affected by d-axis power control, PLL, and DC boost. The above can be discussed as follows:
\vspace{-0.2em} 
\begin{itemize}
    \item \textbf{Equivalent inertia $\boldsymbol{H}^{F}$}: The change rates of the two equivalent frequency components $\omega^{F,Id}$ and $\omega^{F,Pll}$ are each mitigated by their respective equivalent inertia terms $H^{F,Id}$ and $H^{F,Pll}$. Since the total equivalent frequency $\omega^{F}$ is a weighted sum of these components in (\ref{eq:SSDynW_w}), the total equivalent inertia is the weighted superposition of the two equivalent inertia in a parallel configuration:
    \vspace{-0.5em} 
    \begin{equation}\label{eq:HgflSum}
        H^{F} = 1 / \left( c^{Pi}/H^{F,Id} + 1/H^{F,Pll} \right).
        \vspace{-0.7em} 
    \end{equation}
    Unlike the mechanical inertia constant of SGs, the equivalent inertia of GFLs is determined by control parameters as well as the GFL's operating point, such as the levels of active and reactive power.

    \item \textbf{Equivalent governor $\boldsymbol{J^{F}\left(s\right)}$}: For the dynamics of $\omega^{F,Pll}$, in addition to the equivalent inertia $H^{F,Pll}$, higher-order components exist, denoted by $J^{F,Pll}\left(s\right)$. These terms alter the unbalanced power of GFL via $P^{F,Mac,Pll}$, driving $\omega^{F,Pll}$ over a relatively long timescale. Additionally, since $\lim_{s \rightarrow \infty} J^{F,Pll}\left(s\right) = 0$, $\omega^{F,Pll}$ does not provide a steady-state damping effect. It only introduces oscillation in frequency dynamics with an oscillation frequency solved from:
    \vspace{-0.4em} 
    \begin{equation}\label{eq:OscFreq}
        \omega^{Osc} = \frac{1}{4 \pi} \sqrt{4 b^{Pll,0} - b^{Pll,1} }.
        \vspace{-0.5em} 
    \end{equation}
    
    \item \textbf{Proportion coefficient $\boldsymbol{L}^{F}$}: A branch related to $L^{F,Id} \Delta P^{F,Id}$ and $L^{F,Pll} \Delta P^{F,Ele}$ also contributes to the total equivalent frequency $\omega^{F}$. This term causes a sudden step change in the electrical frequencies near the converter. The total step can be expressed as:
    \vspace{-0.3em} 
    \begin{equation}\label{eq:LgflSum}
        L^{F} = c^{Pi} L^{F,Id} + L^{F,Pll}.
        \vspace{-0.5em} 
    \end{equation}

\end{itemize}
\vspace{-0.2em} 

\subsubsection{Time-Variable and Adjustable Characters} As mentioned above, the inertia and other equivalent parameters are formed by GFL control modules, including the PLL and d-axis power control at the frequency dynamics timescale. These parameters are also influenced by the GFL active power $P^{F,Ele0}$, which depends on natural resources, as well as the reactive power $Q^{F*}$ and terminal voltage setpoints $U^{F*}$. The above indicates that the impact of the GFL on COI frequency is time-varying and adjustable, which is totally different from SGs. Together with the linearization coefficient $c^{Pi}$, the impact relationships between equivalent parameters, operating points, and adjustable parameters are summarized in Table \ref{tab:ParaImpc}.

\begin{table}[!t]
	\setlength\tabcolsep{3.0pt}
	\setlength{\aboverulesep}{0.0pt}
	\setlength{\belowrulesep}{1.0pt}
	\centering
	\caption{Impact factors to GFL local dynamics}
    \vspace{-1.0em}
    \begin{tabular}{ccccccccc}
		\toprule
		\specialrule{0em}{0.4pt}{0.4pt}
		\toprule
        \multicolumn{2}{c}{Item} & $P^{F,Ele0}$ & $Q^{F*}$     & $U^{F*}$     & $K_{p}^{Pll}$ & $K_{i}^{Pll}$ & $K_{p}^{Dc}$ & $K_{i}^{Dc}$ \\
        \midrule
        \multirow{3}[2]{*}{$\omega^{F,Id}$} & $H^{F,Id}$ &       &       &       &       &       &       & $\surd$ \\
        \cmidrule{2-9}  & $L^{F,Id}$ &       &       &       &       &       & $\surd$ &       \\
        \cmidrule{2-9}  & $c^{Pi}$ & $\surd$ & $\surd$ &       &       &       &       &       \\
        \midrule
        \multirow{3}[2]{*}{$\omega^{F,Pll}$} & $H^{F,Pll}$ & $\surd$ & $\surd$ & $\surd$ & $\surd$ & $\surd$ & $\surd$ &       \\
        \cmidrule{2-9}  & $J^{F,Pll} \left(s\right)$ & $\surd ^{*}$ & $\surd ^{*}$ & $\surd$ & $\surd$ & $\surd$ & $\surd$ & $\surd$ \\
        \cmidrule{2-9}  & $L^{F,Pll} $ & $\surd$ & $\surd$ & & $\surd$ &       &       &       \\
        \bottomrule
		\specialrule{0em}{0.4pt}{0.4pt}
        \bottomrule
        \multicolumn{9}{l}{\quad $ *$ does not contribute to oscillation frequency $\omega^{Osc}$}
    \end{tabular}
    \vspace{-2.0em}
	\label{tab:ParaImpc}
\end{table}

\vspace{-1.0em} 
\section{Case Studies\label{Sec:CasStu}}
\vspace{-0.3em} 

The proposed model is tested using a modified WECC 9-Bus system. Generator G$3$ represents a GFL, modeled as $80 \times 2$ MW PMSG wind generators. Generators G$1$ and G$2$ remain as SGs, but their inertia is reduced to simulate a low-inertia power system. The topology used for the COI dynamics analysis are illustrated in Figure 2, and the system parameters are provided in Table 1, all given in Appendix \cite{AddDoc}.



\vspace{-1.1em} 
\subsection{Accuracy Validation\label{SubSec:AccuVali}}
\vspace{-0.3em} 

The proposed model, defined by (\ref{eq:SSDynW})-(\ref{eq:SSDynP}), (\ref{eq:PropCoi}), and (\ref{eq:PGFLs}), is validated through a load step disturbance simulation at bus $09$. The dynamic responses are compared with results from detailed EMT simulations in MATLAB/Simulink and two other prominent methods in this field:
\vspace{-0.2em} 
\begin{itemize}
    \item \textbf{Equivalent-rotor-motion method in \cite{HYuan17, RFu22}}: This method introduces the dynamics of an equivalent internal voltage for the GFL, making it behave similarly to a synchronous generator. The internal voltage is derived from the voltage output of the PWM converter behind the terminal voltage, considering the filter in Fig. \ref{fig:GFLCont} as a reactance $X^{Filter}$, as follows:
    \vspace{-0.5em} 
    \begin{equation}\label{eq:YuanCal}
        E^{Int} \angle \delta^{Int} = j X^{Filter} \dot{I}^{F} + \dot{U}^{F}.
        \vspace{-0.7em} 
    \end{equation}
    The connection between generators in a GFL-dominated system is not proposed in \cite{HYuan17, RFu22}, making it impossible to form a COI dynamics model that includes this GFL modeling. Thus, in the comparative analysis, this model replaces the detailed GFL model in the EMT simulation. An intentional error is introduced in $X^{Filter}$ to evaluate the sensitivity of the results, and the reasoning for this adjustment is elaborated upon later.

    \item \textbf{SFR-based method \cite{QShi18}}: This method is widely used in the industry and assumes that a GFL without virtual frequency regulation does not contribute either equivalent inertia or fast frequency response \cite{LXiong22}.
\end{itemize}
\vspace{-0.2em} 

\begin{figure}[!t]
	\centering
	\includegraphics[width=0.46\textwidth]{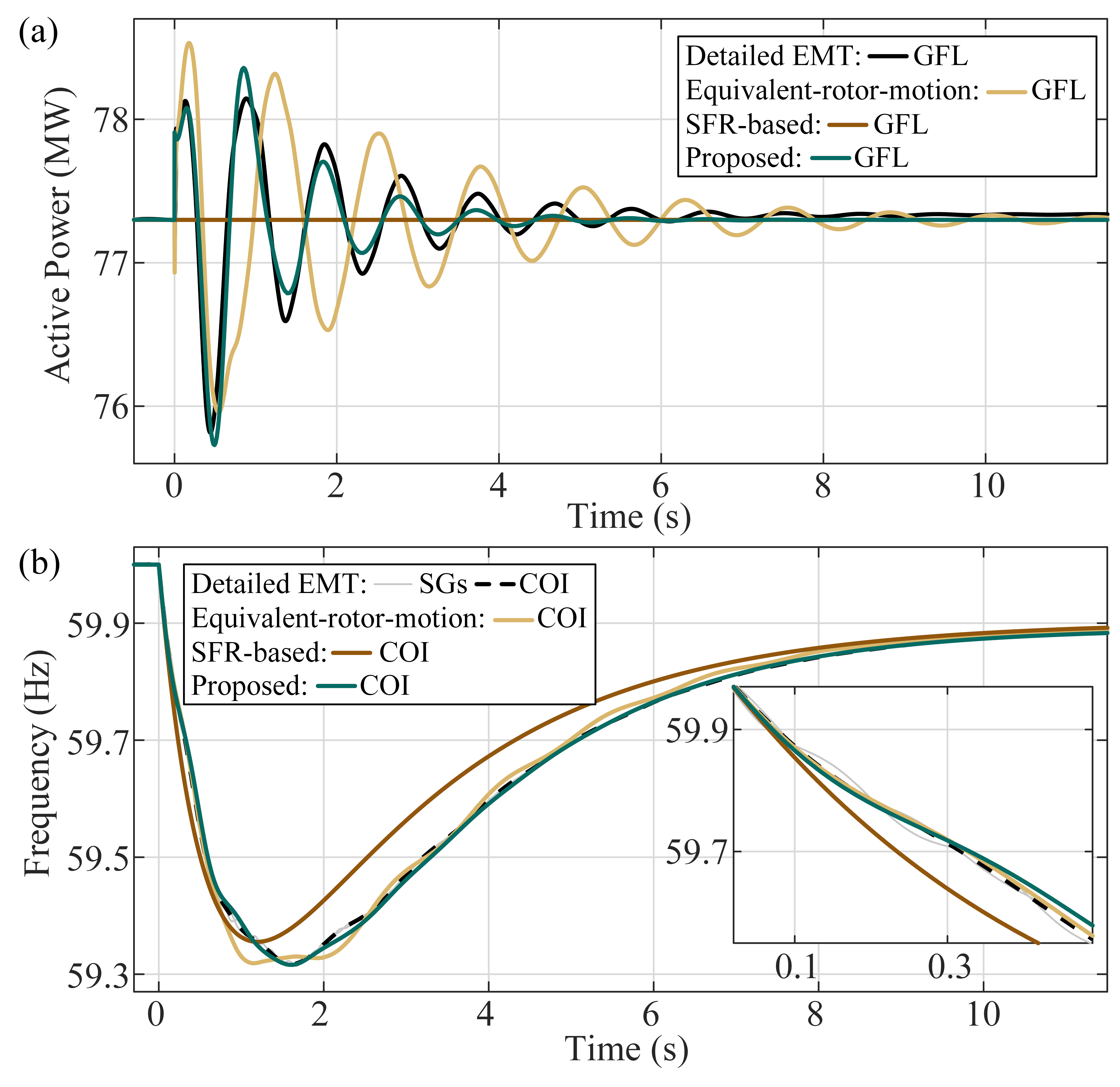}
    \vspace{-1.0em}
	\caption{Model validation with (a) GFL active power and (b) COI frequency.}
    \vspace{-1.45em} 
	\label{fig:cModComp}
\end{figure}

Fig. \ref{fig:cModComp} illustrates the frequency of the SG COI and the active power of the GFL. Using the EMT simulation as the benchmark, the error indices for the other three methods are provided in Table \ref{tab:Error}. The error index for power is calculated as: $ \int \left( P^{cmp} \left( t \right) - P^{emt} \left( t \right) \right) \mathrm{d}t  / \left( \max \left( \Delta P^{emt} \left( t \right) \right) \times T \right)$, in \%, where $T$ is the time window length for error calculation. This index represents the time integral of the power difference between the compared method $P^{cmp}\left( t \right)$ and the EMT simulation $P^{emt}\left( t \right)$, normalized by the maximum deviation of $P^{emt}\left( t \right)$. A similar index is applied for frequency error.

\begin{table}[!t]
	\setlength\tabcolsep{5.8pt}
	\setlength{\aboverulesep}{0.0pt}
	\setlength{\belowrulesep}{1.0pt}
	\centering
	\caption{Model error summary}
    \vspace{-1.0em}
    \begin{tabular}{cccc}
		\toprule
		\specialrule{0em}{0.4pt}{0.4pt}
		\toprule
        Index  & Equivalent-Rotor-Motion & SFR-Based & Proposed \\
        \midrule
        GFL Power & 19.35\% & 11.48\% & 4.41\% \\
        COI Frequency & 1.54\%  & 6.39\%  & 0.56\% \\
        \bottomrule
		\specialrule{0em}{0.4pt}{0.4pt}
        \bottomrule
    \end{tabular}
    \vspace{-2.0em} 
	\label{tab:Error}
\end{table}

For the dynamics obtained from the detailed EMT simulation, the COI frequency represents the average frequency of the two SGs, as depicted in Fig. \ref{fig:cModComp}(b). Following the disturbance, the GFL active power exhibits oscillations, as shown in Fig. \ref{fig:cModComp}(a). These oscillations introduced by the GFL are also reflected in the COI frequency dynamics.

\begin{figure}[!t]
	\centering
	\includegraphics[width=0.46\textwidth]{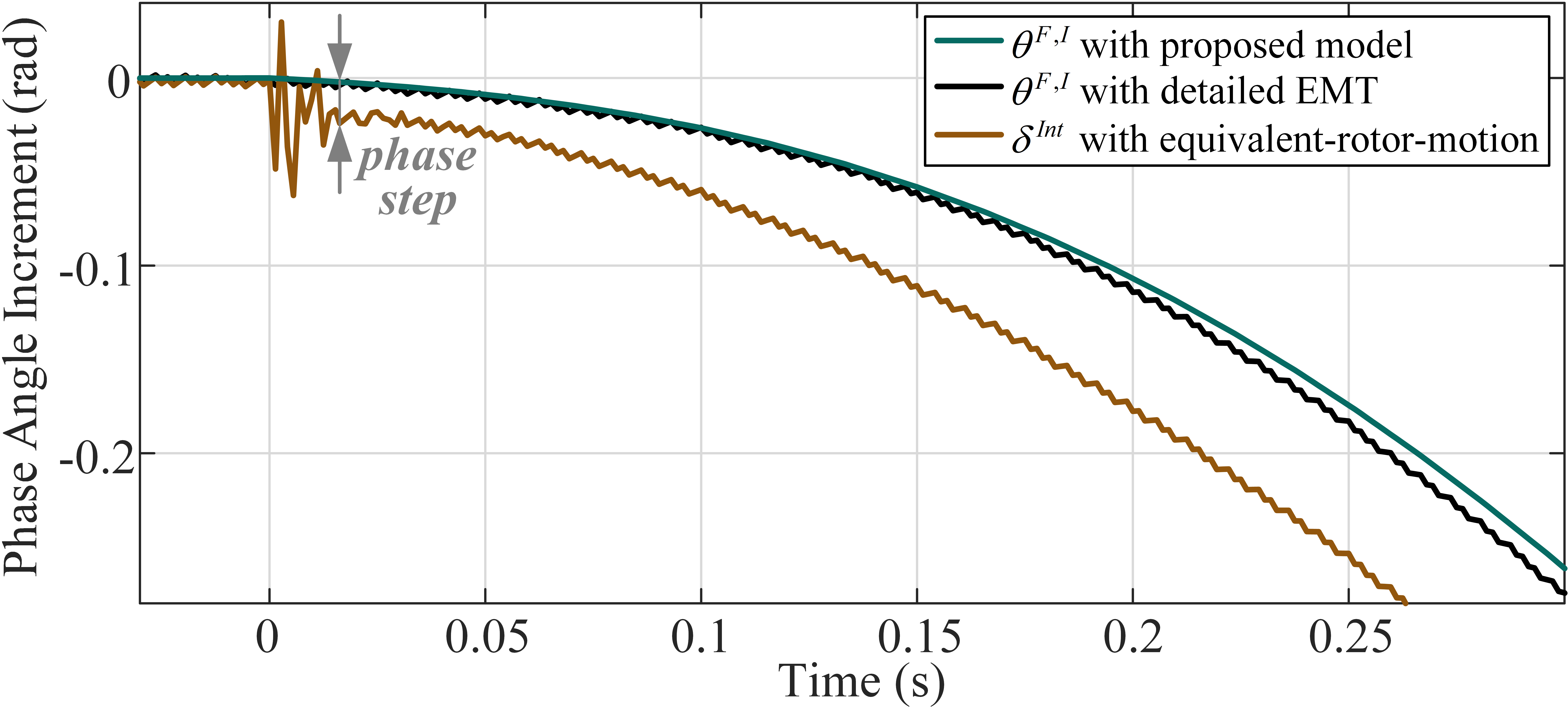}
    \vspace{-1.0em}
	\caption{Phase angle dynamics validation.}
    \vspace{-1.8em}
	\label{fig:cSSComp}
\end{figure}

For the equivalent-rotor-motion model, while the long-term frequency dynamics align closely with the true value in Fig. \ref{fig:cModComp}(b), as the long-term behavior is predominantly governed by the SG-side models, the short-term oscillations in the COI frequency deviate significantly. The reason for this deviation lies in the definition of the internal voltage in (\ref{eq:YuanCal}). In this model, the parameters and structure of the filter, which are typically less critical in frequency dynamics analysis, become pivotal determinants of the model dynamics. This introduces natural inaccuracies due to the challenge of precisely modeling and parameterizing the filter. For instance, when an error is introduced into $X^{Filter}$, an entirely different oscillation pattern emerges in Fig. \ref{fig:cModComp}(a), which propagates to the COI frequency dynamics in Fig. \ref{fig:cModComp}(b). These discrepancies result in errors of 19.35\% in power and 1.54\% in frequency. Beyond the accuracy concerns, a more fundamental issue lies in the definition of the internal voltage itself. The internal voltage should serve as an interface between the internal dynamics of the GFL and the system network, functioning as a state variable at the frequency dynamics timescale. This is analogous to the EMF $E^{G}\angle \delta^{G}$ of SGs. However, as indicated in (\ref{eq:YuanCal}), the defined internal voltage is directly tied to the terminal voltage $\dot{U}^{F}$, which is not a state variable. This linkage makes frequency dynamics analysis problematic. For example, Fig. \ref{fig:cSSComp} illustrates the phase angle dynamics, showing a phase step at the frequency dynamics timescale. This discontinuity hinders effective analysis of the angle interaction between the GFL and the internal angle of the SG.

For the SFR-based model, the GFL dynamics are entirely ignored, resulting in constant active power, as depicted in Fig. \ref{fig:cModComp}(a). This leads to a power error of 11.48\%. Note that this error is lower than that of the equivalent-rotor-motion method, which exhibits an opposite oscillation trend. The COI dynamics in Fig. \ref{fig:cModComp}(b) show an error of 6.39\%. This arises from the high level of simplification in the SFR model

For the proposed model, the power and frequency dynamics presented in Fig. \ref{fig:cModComp} closely align with those obtained from EMT simulations, achieving errors of 4.41\% in power and 0.56\% in frequency. While the improvement in COI frequency error from 1.54\% (using the equivalent-rotor-motion method) to 0.56\% may appear modest, it is notable that half of the error is effectively reduced, and the proposed model successfully corrects the entirely inaccurate GFL power dynamics observed with the equivalent-rotor-motion method. In Fig. \ref{fig:cSSComp}, the angle $\theta^{F,I}$ of the interface state variable does not exhibit a step change after the disturbance, remaining consistent with the $\theta^{F,I}$ dynamics obtained from the EMT simulation.

\vspace{-1.5em} 
\subsection{Revisit COI Frequency Dynamics\label{SubSec:CaseCOI}}
\vspace{-0.3em} 

\begin{figure}[!t]
	\centering
	\includegraphics[width=0.46\textwidth]{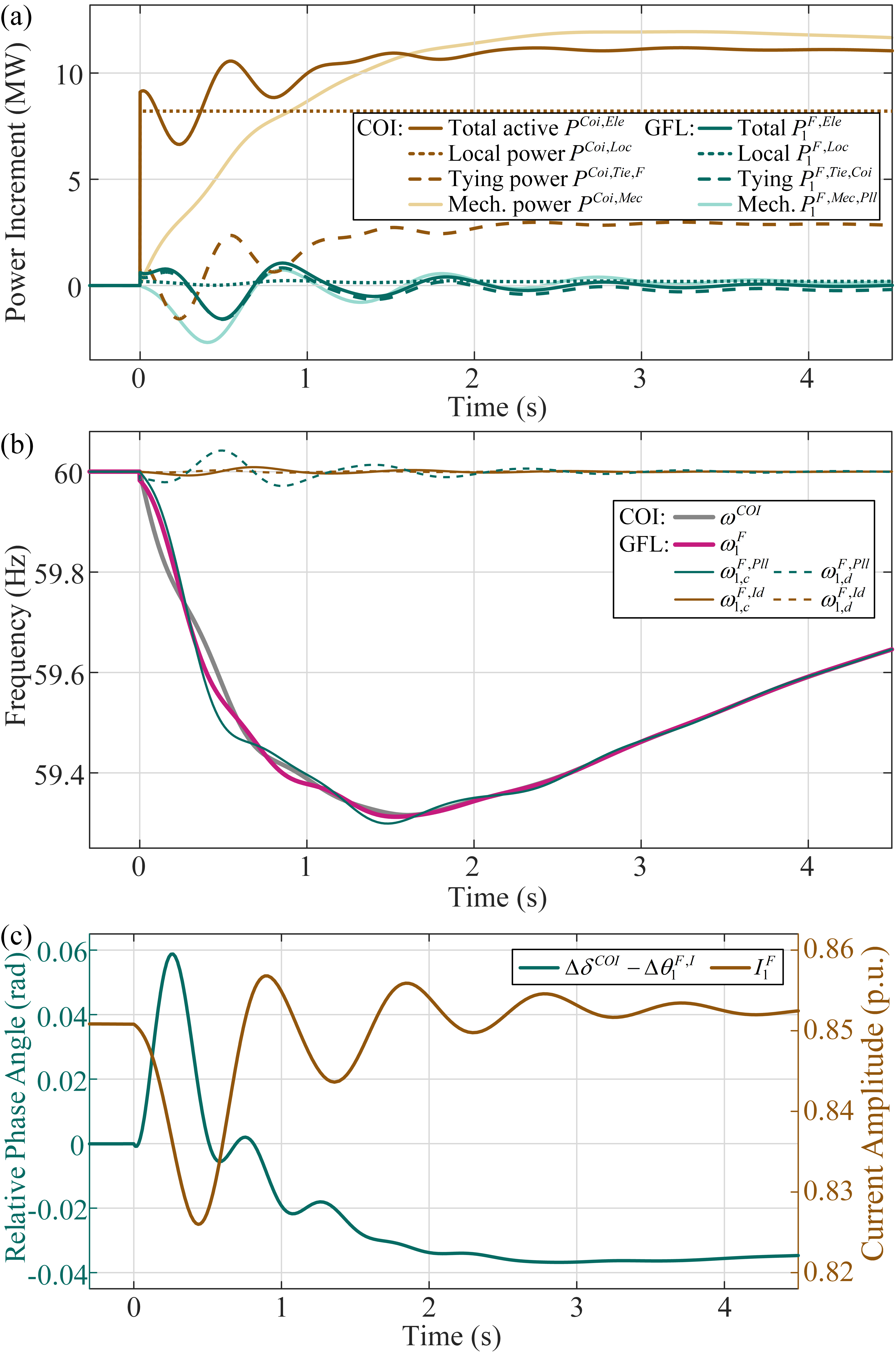}
    \vspace{-1.0em}
	\caption{Dynamics of (a) power, (b) frequency, and (c) state variable.}
    \vspace{-2.22em} 
	\label{fig:cBaseCase}
\end{figure}

The parameters derived from equations (\ref{eq:SSDynW})–(\ref{eq:SSDynP}), (\ref{eq:PropCoi}), and (\ref{eq:PGFLs}) under the system settings provided in Appendix \cite{AddDoc} are summarized in Table \ref{tab:ModPara}. Post-disturbance dynamics, including power, frequency, and interfacing state variables of the GFL, are depicted in Fig. \ref{fig:cBaseCase}. The angle dynamics are represented as relative angles between the GFL phase angle and the COI phase angle. Notably, as the system includes only one GFL, all GFL-related subscripts refer specifically to the $1$st GFL.

\begin{table}[!t]
	\setlength\tabcolsep{4.2pt}
	\setlength{\aboverulesep}{0.0pt}
	\setlength{\belowrulesep}{1.0pt}
	\centering
	\caption{Proposed model with numerical values}
    \vspace{-1.0em}
    \begin{tabular}{cccc}
		\toprule
		\specialrule{0em}{0.4pt}{0.4pt}
		\toprule
        \multicolumn{2}{c}{Model Parts} & Para. (Unit) & Value \\
        \midrule
        \multicolumn{2}{c}{COI Dynamics in (\ref{eq:PropCOI_H}) and (\ref{eq:PropCOI_d})} & $H^{COI}$ (s) & 1.79 \\
        \midrule
        \multirow{12}[6]{*}{\makecell[c]{GFL Dynamics\\in (\ref{eq:SSDynW})-(\ref{eq:SSDynP})}} & Coefficient in          Total Freq. & $c_{1}^{Pi}$ & -0.87 \\
        \cmidrule{2-4}          & \multirow{3}[2]{*}{Equivalent Inertia} & $H_{1}^{F,Id}$ (s) & -3.36 \\
          &       & $H_{1}^{F,Pll}$ (s) & 0.25 \\
          &       & $H_{1}^{F}$ (s) & 0.23 \\
        \cmidrule{2-4}          & \multirow{3}[1]{*}{Proportional Coefficient} & $L_{1}^{F,Id}$ (pu) & 0.006 \\
          &       & $L_{1}^{F,Pll}$ (pu) & -0.080 \\
          &       & $L_{1}^{F}$ (pu) & -0.085 \\
        \cmidrule{2-4}          & \multirow{5}[1]{*}{Equivalent Governor $J_{1}^{F,Pll}\left(s\right)$} & $a_{1}^{Pll,2}$ (pu) & 2.08 \\
          &       & $a_{1}^{Pll,1}$ (pu) & 25.15 \\
          &       & $b_{1}^{Pll,1}$ (pu) & 4.24 \\
          &       & $b_{1}^{Pll,0}$ (pu) & 51.19 \\
          &       & $\omega_{1}^{Osc}$ (Hz) & 1.13 \\
        \midrule
        \multirow{5}[3]{*}{\makecell[c]{Network\\in (\ref{eq:PropCOI_E})-(\ref{eq:PropCOI_P})\\And (\ref{eq:PGFLs})}} & Local Power of COI & $G^{eq\prime}$ (pu) & 2.67 \\
        \cmidrule{2-4}          & \multirow{2}[2]{*}{Tying Power b/w COI-GFL} & $V_{1}^{eq\prime}$ (pu) & -0.84 \\
          &       & $W_{1}^{eq\prime}$ (pu) & 0.23 \\
        \cmidrule{2-4}          & \multirow{2}[2]{*}{Tying Power b/w GFLs} & $R_{1,1}^{eq\prime}$ (pu) & 0.04 \\
          &       & $X_{1,1}^{eq\prime}$ (pu) & 0.18 \\
        \bottomrule
		\specialrule{0em}{0.4pt}{0.4pt}
        \bottomrule
    \end{tabular}
    \vspace{-1.42em} 
	\label{tab:ModPara}
\end{table}

\subsubsection{Tying Power Between COI and GFL}

The disturbance power is primarily absorbed by the COI, as reflected by a step change in the local power $P^{Coi,Loc}$ in Fig. \ref{fig:cBaseCase}(a). This change triggers the post-disturbance dynamics. The interaction between the COI frequency $\omega^{Coi}$ and the GFL equivalent frequency $\omega_{1}^{F}$ is evident in Fig. \ref{fig:cBaseCase}(b). This interaction arises due to the tying power between the COI and the GFL, represented by $P^{Coi,Tie,F}$ in (\ref{eq:PropCOI_P}) and $P_{1}^{F,Tie,Coi}$ in (\ref{eq:PGFLs_T}). Among the two mechanisms of interaction via the GFL angle $\theta_{1}^{F,I}$ and the GFL amplitude $I_{1}^{F}$, the angle contribution is more pronounced. This is reflected in the significant variation from the steady-state operating point, as shown in Fig. \ref{fig:cBaseCase}(c).

In Table \ref{tab:ModPara}, the parameters of the equivalent tie between the COI and the GFL are given as $V_{1}^{eq\prime} = -0.84$ pu and $W_{1}^{eq\prime} = 0.23$ pu. The higher magnitude of $V_{1}^{eq\prime}$ aligns with the discussion in Section \ref{SubSec:SGvsGFL}, confirming that GFLs cannot be aggregated into the SG COI. Despite this, when compared to the high values of system admittance elements, such as $G^{eq\prime}=2.67$ pu, the connection between the COI and the GFL remains relatively weak.

On the GFL side, there exists a local power component, $P^{F,Loc}$, which is influenced by $R_{1,1}^{eq\prime}$ and $X_{1,1}^{eq\prime}$, both of which have lower values in Table \ref{tab:ModPara}. As demonstrated in Fig. \ref{fig:cBaseCase}(a), $P_{1}^{F,Loc}$ has a minimal contribution to the overall dynamics.

\renewcommand{\thefigure}{12}
\begin{figure*}[!t]
	\centering
	\includegraphics[width=0.98\textwidth]{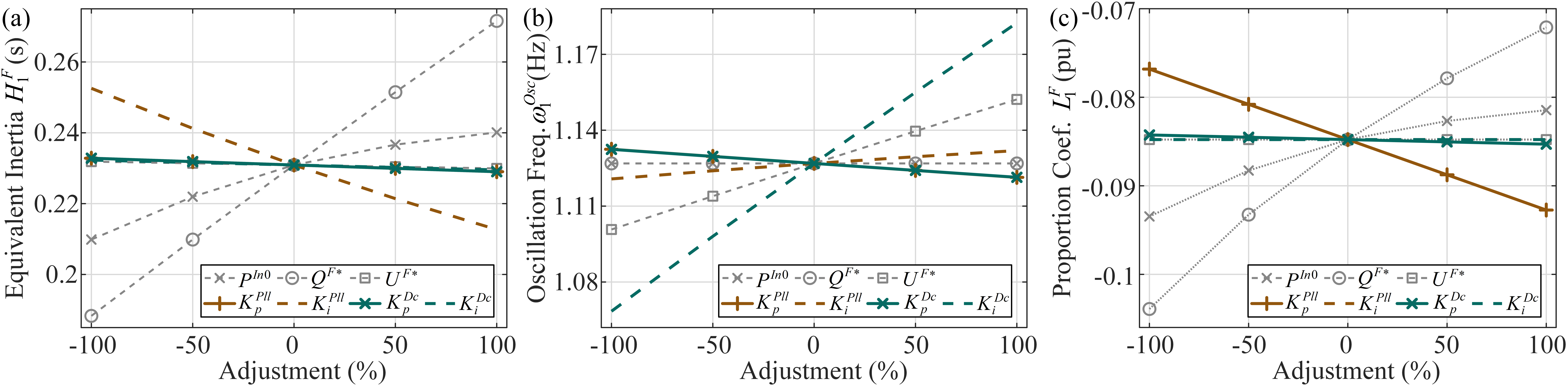}
    \vspace{-1.0em} 
	\caption{Summary of time-variable and adjustable characters for (a) GFL inertia, (b) equivalent generator, and (c) proportional coefficient.}
    \vspace{-1.8em}
	\label{fig:cVar}
\end{figure*}

\subsubsection{Dynamics at GFL Side}

The two components of the GFL equivalent frequency, depicted in Fig. \ref{fig:cBaseCase}(b), collectively constitute the total $\omega^{F}$. Among these, the $\omega_{1}^{F,Pll}$ component dominates, with a more substantial contribution compared to $\omega_{1}^{F,Id}$. Both components are governed by their respective equivalent inertia, governor, and proportional coefficient, which are analyzed separately as follows:

The equivalent inertia $H_{1}^{F,Pll}$ and $H_{1}^{F,Id}$ that governs the continuous components $\omega_{1,c}^{F,Pll}$ and $\omega_{1,c}^{F,Id}$ are $0.25$ s and $-3.36$ s, respectively. The greater magnitude of $H_{1}^{F,Id}$ results in a stronger suppression effect on $\omega_{1,c}^{F,Id}$. It is important to note that the negative value of $H_{1}^{F,Id}$ does not imply negative inertia. Instead, the inertia effect on $\omega_{1,c}^{F,Id}$ in (\ref{eq:SSDynW_w}) is modulated by the coefficient $c_{Pi}$, which is also negative. The combined equivalent inertia, $H_{1}^{F}$, is $0.23$ s, as calculated using (\ref{eq:HgflSum}). This combined inertia is smaller than either $H^{Pll}$ and $H^{Id}$ due to the parallel configuration of $\omega_{eq,c}^{Pll}$ and $\omega_{eq,c}^{Id}$.

The $\omega_{1,c}^{F,Pll}$ component is influenced by the equivalent governor $J_{1}^{F,Pll}\left(s\right)$. Unlike the SG governor, the GFL equivalent governor does not provide steady-state support. This distinction is evident from the fact that $P_{1}^{F,Mac,Pll}$ reaches zero in steady-state following the dynamics, while $P^{Coi,Mec}$ of the COI does not. Additionally, an oscillation frequency $\omega_{1}^{Osc}$ of $1.13$ Hz is observed in the GFL power and frequency.

The proportional coefficient $L_{1}^{F,Pll}=-0.080$ pu is significantly larger than $L_{1}^{F,Id} =-0.006$ pu. This discrepancy is reflected in the substantial frequency step at the disturbance moment in the discontinuous component of the PLL-related frequency $\omega_{1,d}^{F,Pll}$ and, consequently, in the total equivalent frequency $\omega_{1}^{F}$. This starkly contrasts with the COI frequency $\omega^{COI}$, which is constrained by the mechanical rotor dynamics of the SG and, therefore, cannot exhibit a step change.

\subsubsection{Time-Variable and Adjustable Dynamics}

Based on the settings in Appendix \cite{AddDoc}, the frequency dynamics when decreasing $k_{i}^{Pll}$ from the original value of $140$ to $70$ and then to $14$ is shown in Fig. \ref{fig:cPLL}. Under these conditions, the control speed of the PLL is reduced, resulting in an increase in the total equivalent inertia $H_{1}^{F}$ from $0.231$ s to $0.452$ s and then to $1.373$ s. As seen in Fig. \ref{fig:cPLL}, the increased inertia leads to a lower initial RoCoF and a higher nadir in the frequency of the GFL. The COI frequency shows an unchanged initial RoCoF but a higher nadir due to interaction with the GFL frequency through the tying power. It is important to note that the COI frequency dynamics without considering the dynamics of the GFL, representing an SG-dominated system, are also shown. With the initial setting of $k_{i}^{Pll}=140$, the COI frequency nadir is lower than in the SG-dominated case, indicating a worse response. However, as the $k_{i}^{Pll}$ setting is reduced, the COI frequency nadir improves, eventually rising above the nadir of the SG-dominated case. This suggests that reducing the control speed of the integral control in the PLL has a beneficial effect on COI frequency dynamics.

\renewcommand{\thefigure}{10}
\begin{figure}[!t]
	\centering
	\includegraphics[width=0.46\textwidth]{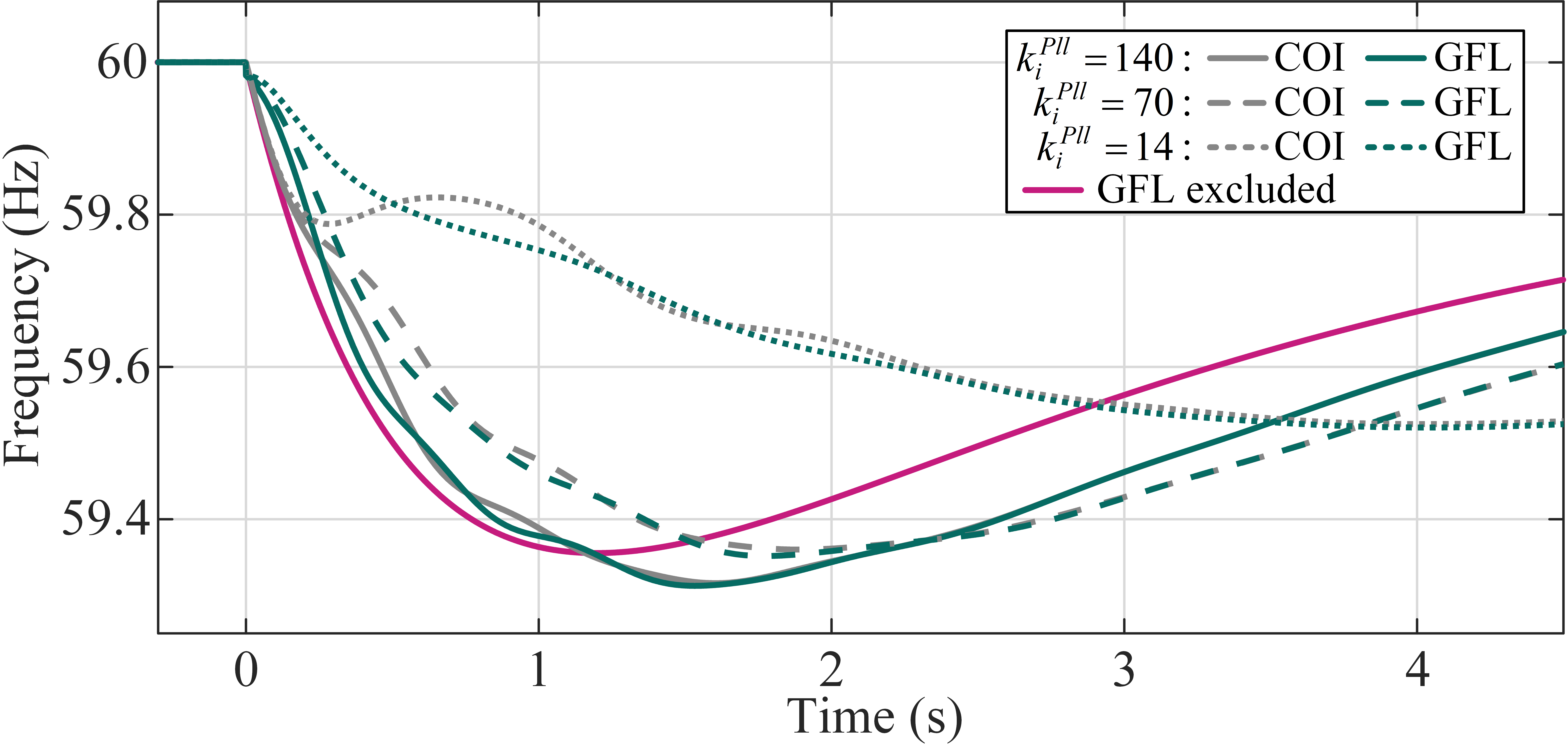}
    \vspace{-1.0em}
	\caption{Frequency trend when reducing PLL integral control setting.}
    \vspace{-1.0em}
	\label{fig:cPLL}
\end{figure}

Another case, where both the proportional and integral control speed of the d-axis power control is reduced, is shown in Fig. \ref{fig:cDC}. In this scenario, the oscillation frequency $\omega_{1}^{Osc}$ of the equivalent generator of the GFL varies significantly. As $k_{p}^{Dc}$/$k_{i}^{Dc}$ and $k_{i}^{Dc}$ are reduced from $0.11$/$2.75$ to $0.011$/$0.275$ and then to $0.0011$/$0.0275$, $\omega_{1}^{Osc}$ decreases from $1.13$ Hz to $0.37$ Hz and then to $0.12$ Hz. As shown in Fig. \ref{fig:cDC}, this reduction results in wider frequency oscillations in both the COI and GFL frequencies, leading to sustained oscillations even as the system frequency recovers from the nadir.

\renewcommand{\thefigure}{11}
\begin{figure}[!t]
	\centering
	\includegraphics[width=0.46\textwidth]{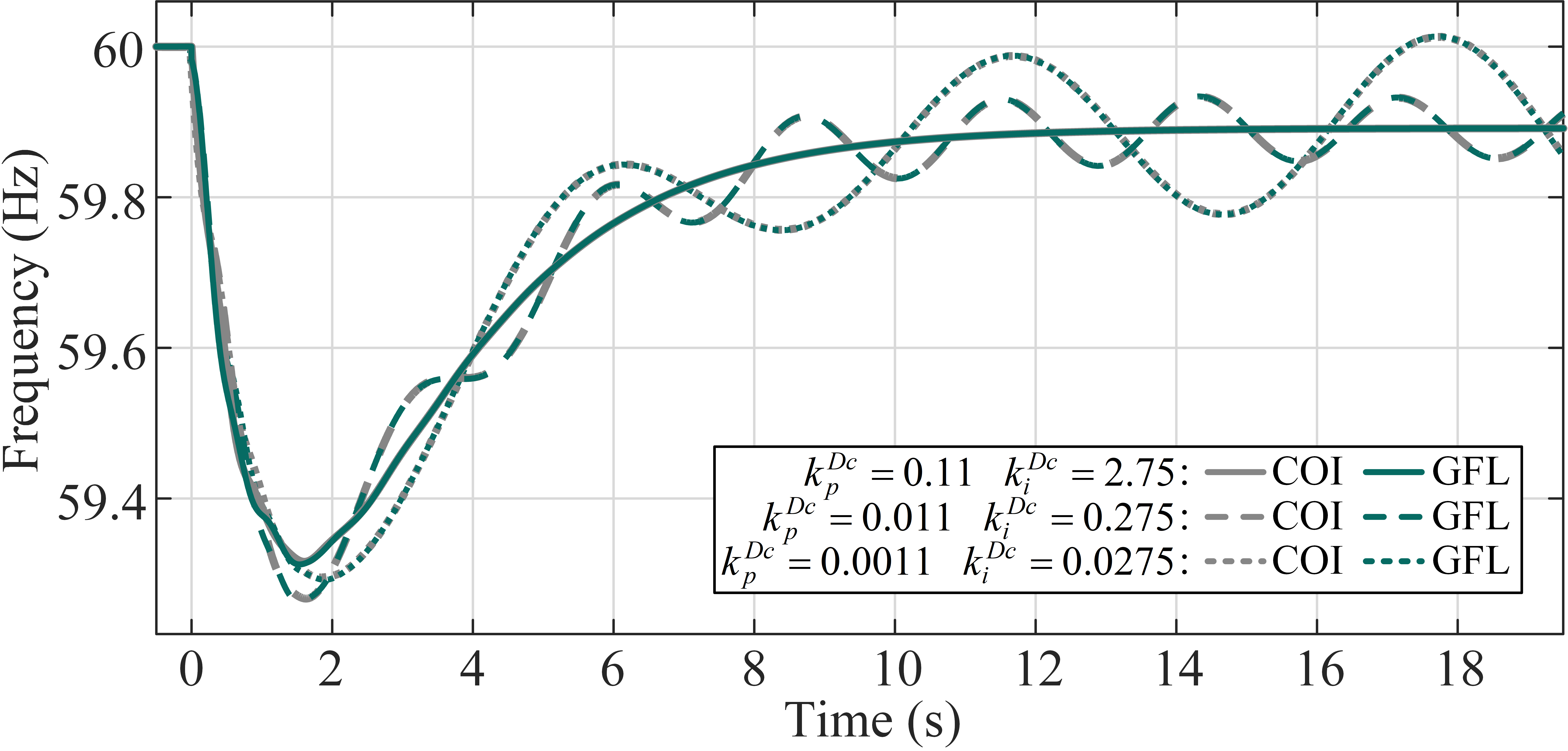}
    \vspace{-1.0em}
	\caption{Frequency trend when reducing proportional and integral control settings of d-axis power control.}
    \vspace{-1.5em} 
	\label{fig:cDC}
\end{figure}

A summary of the time-variable and adjustable dynamics under different GFL conditions is presented in Fig. \ref{fig:cVar}. Figs. \ref{fig:cVar}(a), (b), and (c) illustrate the equivalent inertia, equivalent generator, and proportional coefficient, respectively, with the equivalent generator represented in terms of its oscillation frequency $\omega_{1}^{Osc}$. All adjustable settings are modified based on the baseline parameters listed in Appendix \cite{AddDoc}: the power input $P^{In0}$ varies from $0.238$ pu to $0.857$ pu, the reactive power setting from $0.02$ pu to $0.38$ pu, and the voltage setting from $0.95$ pu to $1.05$ pu. Additionally, $k_{p}^{Pll}$ ranges from $5.4$ to $6.6$, $k_{i}^{Pll}$ from $126$ to $154$, $k_{p}^{Dc}$ from $0.099$ to $0.121$, and $k_{i}^{Dc}$ from $2.475$ to $3.025$. Fig. \ref{fig:cVar}(a) shows that, in addition to the PLL integral gain $k_{i}^{Pll}$, the reactive power and voltage settings significantly influence the GFL equivalent inertia. Other parameters have a lesser effect. Fig. \ref{fig:cVar}(b) highlights the impact of DC control on $\omega_{1}^{Osc}$, along with the effects of voltage settings and PLL control. The active power and reactive power levels have minimal influence. The impact factor of $L_{1}^{F}$ in Fig. \ref{fig:cVar}(c) is analyzed similarly.

\vspace{-1.0em} 
\section{Conclusion\label{Sec:Conclu}}
\vspace{-0.5em} 

Part I of this series comprehensively clarifies the effect of GFLs on system COI frequency dynamics. A multi-generator model of power systems with GFLs is derived. At the generator side, a swing-equation-like GFL model describing the dynamics of equivalent frequency and interface state variables is developed. At the network side, the tying power between the state variables of SGs and GFLs is derived based on the system admittance matrix. Then, it is found that the tying power between SGs is equal in magnitude but opposite in direction, enabling their aggregation into the COI frame. Conversely, the tying power between SGs and GFLs is equal in magnitude but in the same direction, preventing GFLs from being aggregated with SGs. Instead, the equivalent frequency of GFLs interacts with the SG COI frequency through a virtual tie line. Simulations reveal errors in existing approaches for simulating COI frequency dynamics. The connection between SGs and GFLs is shown to be weaker compared to SG connection. The equivalent inertia of GFLs can support the COI inertia through tying power, and the oscillation frequency of the equivalent governor is reflected in the COI dynamics. These effects are time-variable and adjustable, presenting opportunities to compensate for the low-inertia characteristics of GFLs.

\vspace{-1.2em}
\bibliographystyle{IEEEtran}

\bibliography{GFL_Frequency_Dynamics_Part_I}


\end{document}